\documentstyle[amsmath,amssymb,graphicx]{article}

\begin{document}

\thispagestyle{empty}

\baselineskip15pt

\hfill ITEP/TH-12/09

\bigskip

\begin{center}
{\LARGE \bf Introduction to Integral Discriminants}\\
\vspace{0.7cm}
{\large A.Morozov and Sh.Shakirov\footnote{morozov@itep.ru;\ shakirov@itep.ru}}\\
\vspace{0.5cm}
{\em ITEP, Moscow, Russia\\
MIPT, Dolgoprudny, Russia}\\
\end{center}

\bigskip

\centerline{ABSTRACT}

\bigskip

The simplest partition function, associated with homogeneous symmetric forms $S$ of degree $r$ in $n$ variables, is integral discriminant $J_{n|r}\big(S\big) = \int e^{-S(x_1, \ldots, x_n)} \ dx_1 \ldots dx_n$. Actually, $S$-dependence remains the same if $e^{-S}$ in the integrand is substituted by arbitrary function $f(S)$, i.e. integral discriminant is a characteristic of the form $S$ itself, and not of the averaging procedure. The aim of the present paper is to calculate $J_{n|r}$ in a number of non-Gaussian cases. Using Ward identities -- linear differential equations, satisfied by integral discriminants -- we calculate $J_{2|3}, J_{2|4}, J_{2|5}$ and $J_{3|3}$. In all these examples, integral discriminant appears to be a generalized hypergeometric function. It depends on several $SL(n)$ invariants of $S$, with essential singularities controlled by the ordinary algebraic discriminant of $S$.

\tableofcontents

\section{Introduction}

Averaging with exponential weight

$$ \left< \phi \right> = \int \phi(x) e^{-S(x)} d x $$
\smallskip\\
is an important operation in statistical and quantum physics. Function $S(x)$, which determines the weight, is called \emph{action}. The integration domain of $x$-variables is a linear space, which is usually infinite dimensional in real physical applications: for example, a space of paths in quantum mechanics or a space of field configurations in quantum field theory. Infinite dimension of the space introduces additional complications: the integral is not always well-defined. For this reason, it is important first to study such integrals in finite dimension $n$. After this is done, one can take a $n \rightarrow \infty$ limit.

However, even in finite dimension $n$, the averaging operation is not yet fully understood. Most studied are \emph{Gaussian} integrals: that is, when $S$ is quadratic in $x$-variables. The simplest Gaussian integral

$$ \left< 1 \right> = \int e^{- S_{ij} x_{i} x_{j}} d^n x $$
\smallskip\\
is easily calculated and expressed through an invariant quantity -- determinant -- e.g. by diagonalising $S$:

$$ \int e^{- S_{ij} x_{i} x_{j}} d^n x = \int e^{ - \lambda_1 x_{1}^2 - \ldots - \lambda_n x_{n}^2} \ d^n x = \dfrac{1}{\sqrt{ \lambda_{1} \ldots \lambda_{n} }} \int e^{ - x_{1}^2 - \ldots - x_{n}^2 } \ d^n x \sim \dfrac{1}{\sqrt{ \det S }}$$
\smallskip\\
The integral which factors out is just an $S$-independent constant, which can be finite or infinite depending on the contour of integration. Therefore, essential $S$-dependence of this Gaussian integral is given by $(\det S)^{-1/2}$. Any other Gaussian integral -- with non-homogeneous quadratic $S$ or non-trivial $\phi$ -- is equally easy, because we have enough freedom to transform quadratic $S$ to diagonal or any other desired form. The possibility of diagonalisation greatly simplifies calculations with matrices (tensors with two indices).

Unfortunately, such methods do not work when $S$ is cubic or higher degree. This can be seen already from dimension counting. The number of independent coefficients $S_{ijk}$, which is $n(n+1)(n+2)/6$, exceeds the number $n^2 - 1$ of available $SL(n)$ transformations, so it is not generally possible to diagonalize a cubic action -- the group $SL(n)$ is too small. That is why the integral

$$ \int e^{- S_{ijk} x_{i} x_{j} x_{k}} d^n x $$
\smallskip\\
and its higher degree analogues

\begin{align}
J_{n|r} = \int e^{- S(x_1, \ldots, x_n)} d^n x, \ \ \ \ S(x_1, \ldots, x_n) = S_{i_1, \ldots, i_r} x_{i_1} \ldots x_{i_r}
\label{IntDis}
\end{align}
\smallskip\\
despite simply-looking, still remain \emph{terra incognita}. In \cite{Dolotin, nolal} integral $J_{n|r}(S)$ was named the \emph{integral discriminant} of $S$, because in the simplest cases this integral is just a power of algebraic discriminant $D_{n|r}(S)$. For example, there is an inspiring formula for 3-forms in two variables

$$
J_{2|3} = \int e^{ - \big( a x^3 + b x^2 y + c x y^2 + d y^3 \big) } dx dy = \big( 27 a^2 d^2 - b^2 c^2 - 18 a b c d + 4 a c^3 + 4 b^3 d \big)^{-1/6} = \left( D_{2|3}\right)^{-1/6}
$$
\smallskip\\
which shows that $J_{2|3}$ and $D_{2|3}$ are, indeed, related. However, when one goes to higher $n$ and $r$, the relation gets more complicated: $D_{n|r}$ defines only the singularities of $J_{n|r}$, while non-singular behaviour is controlled by other algebraic invariants. Thus, theory of integral discriminants is closely connected to \emph{invariant theory} \cite{invtheory} and can be viewed as one of the branches of \emph{non-linear algebra} \cite{nolal}, \cite{GKZ} - \cite{Ainfty}.

It would be very interesting to find a closed formula for generic $J_{n|r}$, because it could provide exciting new tools in QFT, statistics and other fields where non-Gaussian averaging is used. In this paper, we make a step in this direction and find integral discriminants explicitly for 3-forms, 4-forms and 5-forms in two variables, and for 3-forms in three variables. Our results indicate an intriguing connection between integral discriminants (\ref{IntDis}) and special functions known as generalized hypergeometric functions \cite{genhypergeom}.

\section{Ward identities}

When the action is non-quadratic, diagonalisation and similar linear-algebra tricks fail. To handle non-Gaussian integrals, one needs essentially different methods. One of such methods (actually originated in the context of quantum field theory) is to find a differential equation, satisfied by the integral as a function of its parameters, see \cite{ward,picard} for typical applications and references. If such differential equation exists, we call it Ward identity (even if it is not directly induced by a change of integration variables).

Since Ward identities play the central role in present paper, let us give a pair of simple examples to clarify this issue. An integral

$$ F(a) = \int e^{-x^2/2 + ax} dx $$
\smallskip\\
satisfies a Ward identity

$$\left( a - \dfrac{\partial}{\partial a} \right) F(a) = 0 $$
\smallskip\\
because

$$\left( a - \dfrac{\partial}{\partial a} \right) F(a) = \int (a - x) e^{-x^2/2 + ax} dx = \int \left( \dfrac{\partial}{\partial x} \ e^{-x^2/2 + ax} \right) dx = 0$$
\smallskip\\
In contrast with diagonalisation, this method is perfectly generalisable to non-Gaussian integrals: say,

$$ G(a) = \int e^{-x^3/3 + ax} dx $$
\smallskip\\
satisfies a Ward identity

$$\left( a - \dfrac{\partial^2}{\partial a^2} \right) G(a) = 0 $$
\smallskip\\
because

$$\left( a - \dfrac{\partial^2}{\partial a^2} \right) G(a) = \int (a - x^2) e^{-x^3/3 + ax} dx = \int \left( \dfrac{\partial}{\partial x} \ e^{-x^3/3 + ax} \right) dx = 0$$
\smallskip\\
In this way the problem of non-Gaussian integrals is reduced to another problem -- of differential equations.  This is a much easier problem, especially if differential equations are linear. In the first case we have

$$ a F(a) = \dfrac{\partial}{\partial a} F(a) \ \ \Longrightarrow \ \ F(a) = c \cdot e^{a^2/2} $$
while in the second case

$$ a G(a) = \dfrac{\partial^2}{\partial a^2} G(a) \ \ \Longrightarrow \ \ G(a) = c_1 \cdot \mbox{ Ai }(a) + c_2 \cdot \mbox{ Bi }(a) $$
\smallskip\\
where Ai and Bi are special functions -- Airy functions of the first and second kind. Note that, there is only one linear independent solution in the Gaussian case, while in the non-Gaussian case there are two linear-independent solutions. This is because to correctly define an integral, one still needs to specify an integration contour. Different integration contours provide different solutions of the Ward identity.

This relationship between Ward identities and integration contours is quite important, so let us add more details. As the simplest option, the contour of integration in the integral

\begin{figure}[t]
\begin{center} \includegraphics[width=250pt]{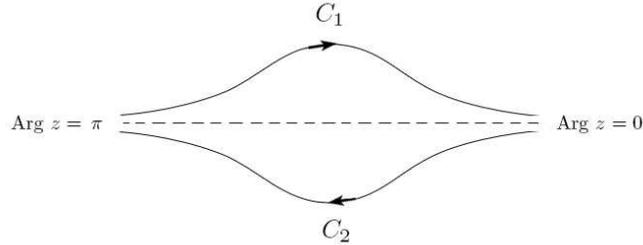}
\caption{Two, out of infinitely many, admissible contours of integration for the integral $\int e^{-x^2/2 + ax} dx$.}
\end{center}
\end{figure}

$$ F(a) = \int\limits_{C} e^{-x^2/2 + ax} dx $$
\smallskip\\
can be chosen as $C = \mbox{real axis}$. However, this choice is by no means unique. In fact, any other contour $C$ which asymptotically tends to the lines $\mbox{Arg } z = 0$ and $\mbox{Arg } z = \pi / 2$, is admissible. A contour $C$ is said to be admissible, if the integral over $C$ converges. A few admissible contours are shown at Fig. 1. Note that, to ensure vanishing of the integral of full derivative (and thus validity of Ward identities) we consider only closed contors -- the contours at Fig. 1. are closed on the Riemann sphere, if the infinitely remote point is taken into account. Since Ward identity in this case is a differential equation of first order, all the contours give one and the same answer $F(a) = \exp \big( a^2/2 \big)$ up to proportionality.

More interesting is the non-Gaussian integral

$$ G(a) = \int\limits_{C} e^{-x^3/3 + ax} dx $$
\smallskip\\
Note that, the real axis is no longer an admissible contour, since $\exp\big( -x^3 \big)$ grows to infinity when $x \rightarrow -\infty$. In this case, admissible is any contour $C$ which asymptotically tends to the lines $\mbox{Arg } z = 0, \ \mbox{Arg } z = 2 \pi / 3$ and $\mbox{Arg } z = 4 \pi / 3$. A few admissible contours are shown at Fig. 2. Again, to ensure vanishing of the integral of full derivative, we consider only closed contors.
Since Ward identity is second order in this case, there are two essentially different integration contours, say, $C_1$ and $C_2$. An integral over arbitrary contour -- the general solution of the Ward identity -- is given by linear combination

$$ \int\limits e^{-x^3/3 + ax} dx \ = \ c_1 \ \int\limits_{C_1} e^{-x^3/3 + ax} dx \ + \ c_2 \ \int\limits_{C_2} e^{-x^3/3 + ax} dx $$
\smallskip\\
To summarize the above examples, Ward identity is the main differential equation which governs all the contours at once. The choice of particular contour corresponds to the choice of particular solution of the Ward identity. For this reason, Ward identities are especially convenient to study of properties, which are invariant under change of integration contour. Additional details can be found in \cite{ward}.

\begin{figure}[t]
\begin{center}
\includegraphics[totalheight=230pt,clip]{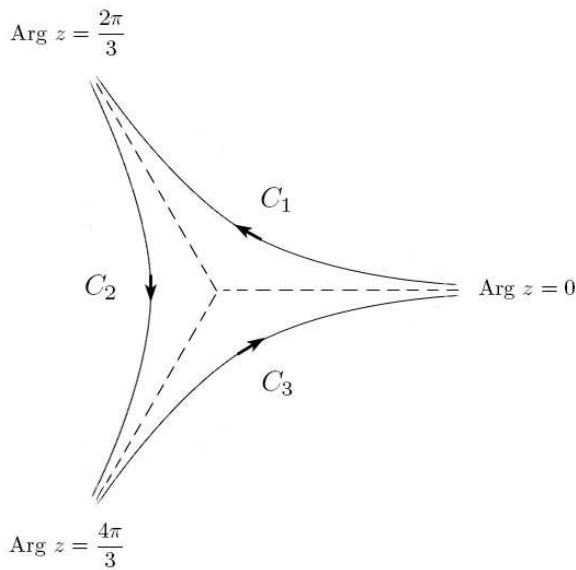}
\caption{Three, out of infinitely many, admissible contours of integration for the integral $\int e^{-x^3/3 + ax} dx$.}
\end{center}
\end{figure}

\section{Integral Discriminants}

\paragraph{Definition.}In this paper we study a specific class of non-Gaussian integrals: \emph{integral discriminants}

$$
 J_{n|r} \big( S \big) = \int e^{-S(x_1,x_2, \ldots, x_n)} dx_1 \ldots dx_n
$$
\smallskip\\
associated with homogeneous symmetric $r$-forms $S(x_1,x_2,\ldots,x_n)$. There are two different notations for symmetric forms, which can be useful under different circumstances: tensor notation

$$ S(x_1,x_2,\ldots,x_n) = \sum\limits_{i_1,i_2, \ldots, i_r = 1}^{n} S_{i_1,i_2, \ldots, i_r} x_{i_1} x_{i_2} \ldots x_{i_r} $$
\smallskip\\
and monomial notation

$$ S(x_1,x_2,\ldots,x_n) = \sum\limits_{a_1 + a_2 + \ldots + a_n = r } s_{a_1,a_2 \ldots a_n } x_{1}^{a_1} x_{2}^{a_2} \ldots x_{n}^{a_n} $$
\smallskip\\
To distinguish between these notations, we denote coefficients by capital and small letters, respectively. Note, that in tensor notation coefficients have $r$ indices, while in monomial notation they have $n$ indices.

\paragraph{The choice of contour.}Being non-Gaussian integrals, integral discriminants of course depend on the choice of integration contour. One has to remember, that not every contour is admissible: for given $S$, only restricted set of contours is allowed. Say, for positive-definite quadratic forms $S$ admissible is any contour, which is asymptotically pure real (see the previous section, especially Fig. 1 and Fig. 2 for simple examples). At the same time for negative-definite quadratic forms admissible is any contour, which is asymptotically pure imaginary. In this paper, we do not describe the contour dependence explicitly. Instead, we concentrate on contour-independent properties of integral discriminants.

\paragraph{Independence on the form of action.}An important feature of integral discriminants, which is due to homogeneity of $S$, is the possibility to substitute the function $e^{-S}$ under the integral with arbitrary function (or, better to say, arbitrary good function) $f(S)$:

\begin{align}
\int e^{- S(x_1, \ldots, x_n)} \ dx_1 \ldots dx_n \sim \int f \Big( S(x_1, \ldots, x_n) \Big) \ dx_1 \ldots dx_n
\label{Prop2}
\end{align}
\smallskip\\
Even before specifying the class of good functions, let us consider a simple illustration with $f(S) = e^{-S^2}$:

$$
\int e^{- \big(S_{ij} x_i x_j\big)} \ d^n x = \int e^{- \big(\lambda_1 x_1^2 + \ldots + \lambda_n x_n^2\big)} \ d^n x = \dfrac{1}{\sqrt{\lambda_1 \ldots \lambda_n}} \int e^{- \big(x_1^2 + \ldots + x_n^2\big)} \ d^n x = \dfrac{{\rm const}}{\sqrt{\det S}}
$$

$$
\int e^{- \big(S_{ij} x_i x_j\big)^2} \ d^n x = \int e^{- \big(\lambda_1 x_1^2 + \ldots + \lambda_n x_n^2\big)^2} \ d^n x = \dfrac{1}{\sqrt{\lambda_1 \ldots \lambda_n}} \int e^{- \big(x_1^2 + \ldots + x_n^2\big)^2} \ d^n x = \dfrac{{\rm const^{\prime}}}{\sqrt{\det S}}
$$
\smallskip\\
To specify the class of good functions and prove (\ref{Prop2}), let us make a change of integration variables

$$x_1 = \rho, \ \ x_2 = \rho z_2, \ \ x_3 = \rho z_3, \ \ \ldots, x_n = \rho z_n $$
\smallskip\\
i.e. pass from homogeneous coordinates $x_i$ to non-homogeneous coordinates $z_i = x_i / x_1$. Then

$$ \int e^{- S(x_1, x_2, \ldots, x_n)} \ dx_1 \ldots dx_n = \int \rho^{n - 1} d\rho \int dz_2 \ldots dz_n \ e^{- \rho^r S(1, z_2, \ldots, z_n)} = $$

$$ = \left( \int \rho^{n - 1} e^{- \rho^r} d\rho \right) \cdot \int \dfrac{dz_2 \ldots dz_n}{S(1, z_2, \ldots, z_n)^{n/r}} $$
\smallskip\\
For the right hand side of (\ref{Prop2}) we have

$$ \int f\Big( S(x_1, x_2, \ldots, x_n) \Big) \ dx_1 \ldots dx_n = \int \rho^{n - 1} d\rho \int dz_2 \ldots dz_n \ f\Big( \rho^r S(1, z_2, \ldots, z_n) \Big) = $$

$$ = \left( \int \rho^{n - 1} f\Big( \rho^r \Big) d\rho \right) \cdot \int \dfrac{dz_2 \ldots dz_n}{S(1, z_2, \ldots, z_n)^{n/r}} $$
\smallskip\\
Both integrals over $\rho$ are just $S$-independent constants. As one can see, (\ref{Prop2}) is valid, iff these integrals

$$ \int \rho^{n - 1} e^{- \rho^r} d\rho \ \ \ \ {\rm and} \ \ \ \ \int \rho^{n - 1} f\Big( \rho^r \Big) d\rho $$
\smallskip\\
are finite over one and the same contour. This condition specifies the class of good functions $f(S)$. For example, all functions $ f(S) = \exp \big( - S^k \big) $ for $k > 0$ fall into this class. Relation (\ref{Prop2}) is therefore proved. As a byproduct, we have obtained a non-homogeneous integral representation

\begin{equation}
\addtolength{\fboxsep}{5pt}
\boxed{
\begin{gathered}
J_{n|r} \big( S \big) = {\rm const} \cdot \int \dfrac{dz_2 \ldots dz_n}{S(1, z_2, \ldots, z_n)^{n/r}}
\end{gathered}
}\label{NonHomogen}
\end{equation}
\smallskip\\
A useful complement to the definiton of integral discriminants, representation (\ref{NonHomogen}) highlights two properties, which were less evident in (\ref{IntDis}): the \emph{scaling dimension}
\begin{align}
 J_{n|r} \big( \lambda S \big) = \lambda^{-n/r} \ J_{n|r} \big( S \big)
\label{Prop1}
\end{align}
and the \emph{vertical symmetry}
\begin{align}
 I_{n|kr} \big( S^k \big) \sim  J_{n|r} \big( S \big)
\label{Vertical}
\end{align}

\paragraph{Ward identities.}We now turn to Ward identities, satisfied by integral discriminants with respect to their parameters. Let us introduce the \emph{correlation functions}

$$ \Big< \phi(x_1, \ldots, x_n) \Big> = \int \phi(x_1, \ldots, x_n) e^{- S(x_1, x_2, \ldots, x_n)} \ dx_1 \ldots dx_n $$
\smallskip\\
These correlation functions satisfy

\begin{align}
\dfrac{\partial}{\partial s_{a_1, \ldots, a_n}} \Big< \phi(x_1, \ldots, x_n) \Big> = - \Big< x_1^{a_1} \ldots x_n^{a_n} \phi(x_1, \ldots, x_n) \Big>
\label{DiffIdentity}
\end{align}
\smallskip\\
and

$$ \dfrac{\partial}{\partial s_{a_1, \ldots, a_n}} \dfrac{\partial}{\partial s_{b_1, \ldots, b_n}}  \Big< \phi(x_1, \ldots, x_n) \Big> = \Big< x_1^{a_1 + b_1} \ldots x_n^{a_n + b_n} \phi(x_1, \ldots, x_n) \Big> $$
\smallskip\\
The right hand side depends only on the sum of indices $a_i + b_i$, not on $a_i$ and $b_i$ separately. For this reason, correlation functions satisfy a system of homogeneous second order differential equations:

\begin{align}
\left( \dfrac{\partial}{\partial s_{\vec a}} \dfrac{\partial}{\partial s_{\vec b}} - \dfrac{\partial}{\partial s_{\vec p}} \dfrac{\partial}{\partial s_{\vec q}} \right) \Big< \phi(x_1, \ldots, x_n) \Big> = 0, \ \ \ \ \ {\vec a} + {\vec b} = {\vec p} + {\vec q}
\label{WardPhi}
\end{align}
\smallskip\\
In particular, if we set $\phi(x_1, \ldots, x_n) = 1$ we find that integral discriminant $\Big< 1 \Big> = J_{n|r}$ satisfies Ward identities

\begin{equation}
\addtolength{\fboxsep}{5pt}
\boxed{
\begin{gathered}
\left( \dfrac{\partial}{\partial s_{\vec a}} \dfrac{\partial}{\partial s_{\vec b}} - \dfrac{\partial}{\partial s_{\vec p}} \dfrac{\partial}{\partial s_{\vec q}} \right)  J_{n|r} \big( S \big) = 0, \ \ \ \ \ {\vec a} + {\vec b} = {\vec p} + {\vec q}
\end{gathered}
}\label{Ward}
\end{equation}
\smallskip\\
These equations do not exhaust the set of all Ward identities: there are more. Notice that the differential operator in the left hand side of (\ref{Ward}) annihilates not only the integral, but also the integrand:

$$
\left( \dfrac{\partial}{\partial s_{\vec a}} \dfrac{\partial}{\partial s_{\vec b}} - \dfrac{\partial}{\partial s_{\vec p}} \dfrac{\partial}{\partial s_{\vec q}} \right) e^{- S(x_1, x_2, \ldots, x_n)} = 0, \ \ \ \ \ {\vec a} + {\vec b} = {\vec p} + {\vec q}
$$
\smallskip\\
and even

$$
\left( \dfrac{\partial}{\partial s_{\vec a}} \dfrac{\partial}{\partial s_{\vec b}} - \dfrac{\partial}{\partial s_{\vec p}} \dfrac{\partial}{\partial s_{\vec q}} \right) f \Big( S(x_1, \ldots, x_n) \Big) = 0, \ \ \ \ \ {\vec a} + {\vec b} = {\vec p} + {\vec q}
$$
\smallskip\\
which justifies (\ref{Prop2}) once again. Usually, Ward differential operators do not annihilate the integrand, they just transform the integrand into a full derivative (which implies vanishing of the integral over any closed contour). Differential operator in the left hand side of  (\ref{Ward}) is therefore too special; one can expect other Ward identities to exist. To find them, let us consider the most general vanishing correlator -- an integral of full derivative

$$ \int \dfrac{\partial}{\partial x_i} \left( \phi(x_1, \ldots, x_n) e^{- S(x_1, x_2, \ldots, x_n)} \right) \ dx_1 \ldots dx_n  = 0 $$
\smallskip\\
Taking the derivative, we obtain Ward identities for correlation functions:

\begin{align}
\Big< \dfrac{\partial \phi(x_1, \ldots, x_n)}{\partial x_i} \Big> = \Big< \phi(x_1, \ldots, x_n) \dfrac{\partial S(x_1, \ldots, x_n)}{\partial x_i} \Big>
\label{GeneralWard}
\end{align}
\smallskip\\
For now we are not interested in all correlation functions, only in the integral discriminant. We need to rewrite the equations (\ref{GeneralWard}) as differential equations on $J_{n|r}$. There are several ways to do this, the simplest way is to set $\phi(x_1, \ldots, x_n) = x_j$ and use the identity (\ref{DiffIdentity}) to get rid of remaining correlators. Doing so, we obtain equations

\begin{equation}
\addtolength{\fboxsep}{5pt}
\boxed{
\begin{gathered}
\left( \delta_{ij} + a_i \ {\hat A}_{ij} \right) J_{n|r} = 0
\end{gathered}
}\label{GeneralVaccuumWard}
\end{equation}
\smallskip\\
where operators

\begin{align}
 {\hat A}_{ij} = \sum\limits_{a_1 + a_2 + \ldots + a_n = r } s_{a_1 \ldots a_{i} - 1 \ldots a_{j} + 1 \ldots a_n} \dfrac{\partial}{\partial s_{a_1 \ldots a_n}}
\label{GLn}
\end{align}
\smallskip\\
form the $GL(n)$ algrebra:

$${\hat A}_{ij} {\hat A}_{kl} - {\hat A}_{kl} {\hat A}_{ij} = {\hat A}_{il} \delta_{kj} - {\hat A}_{kj} \delta_{il}$$
\smallskip\\
Note also, that other choices of $\phi(x_1, \ldots, x_n)$ in (\ref{GeneralWard}) do not give new Ward identities: everything, that can be obtained in this way, will be equivalent to (\ref{GeneralVaccuumWard}). We conclude that the complete set of Ward identities for $J_{n|r}$ consists of equations (\ref{Ward}) and (\ref{GeneralVaccuumWard}).

\paragraph{$SL(n)$ invariance.} In solving these Ward identities we start from (\ref{GeneralVaccuumWard}), because they are first order. In fact, equations (\ref{GeneralVaccuumWard}) simply reflect the $GL(n)$-covariance (= $SL(n)$ invariance + correct scaling) of the model. This becomes clear, if we separate the generators of $GL(n)$ into two parts: the dilatation (degree) operator

$${\hat A}_{11} + \ldots + {\hat A}_{nn}$$
\smallskip\\
and the other $n^2 - 1$ operators, which form a representation of $SL(n)$. The first relation implies

$$\left( {\hat A}_{11} + \ldots + {\hat A}_{nn} \right) J_{n|r} = -\dfrac{n}{r} \ J_{n|r} $$
\smallskip\\
which is nothing but the scaling dimension (\ref{Prop1}). The other $n^2 - 1$ relations imply, that $J_{n|r}$ is $SL(n)$-invariant function of $S$. Thus, equations (\ref{GeneralVaccuumWard}) are completely solved by any $SL(n)$-invariant function of $S$, which is in addition homogeneous in $S$ of degree $-n/r$. A natural question is: is there any simple description of such functions? Actually, the answer to this question is positive: any $SL(n)$-invariant function can be uniqely represented as a function of the elementary invariants $I_{k}$

$$ J_{n|r} = F\big\{ I_k \big\} $$
\smallskip\\
much in the same way as any $SL(n)$-invariant function of a matrix $A^{i}_{j}$ can be uniquely represented as a function of elementary invariants ${\rm tr} A^k$. Unfortunately, in the case of symmetric tensors $S_{i_1, \ldots, i_r}$ classification of these elementary invariants $I_{k}$ is not that easy, as in the case of matrices. Since a symmetric tensor with $r$ indices is much more complicated, than a matrix with two indices, no explicit formula like $I_k = {\rm tr} A^k$ is available. The study of properties of the elementary invariants $I_{k}$, of different kinds of explicit formulas and relations between them, is a classical branch of science known as \emph{invariant theory} \cite{invtheory}.

\paragraph{The number of $SL(n)$ invariants.} Given parameters $n$ and $r$ of the form $S$, one can easily find the number of elementary invariants $I_{k}$, of which all other invariants are various functions. Indeed, the linear space of forms $S$ of type $n|r$ has dimension

$$ \dim S_{n|r} = \dfrac{(n + r - 1)!}{r! (n-1)!} $$
\smallskip\\
that is the number of independent coefficients of symmetric tensor $S_{i_1 \ldots i_r}$. The group $SL(n)$ acts on this space, dividing it into orbits. All forms, connected by $SL(n)$ transformations $ S_{i_1 \ldots i_r}  \mapsto S_{i_1 \ldots i_r} U^{i_1}_{j_1} \ldots U^{i_n}_{j_n}, \ \ \ U \in SL(n) $ belong to one orbit. Ward identities imply, that $J_{n|r}$ does not depend on coordinates along the orbit ("angular" variables). It depends only on transverse coordinates which label orbits ("radial" variables). Simple counting of dimensions implies, that dimension of the space of orbits, i.e. the number of radial coordinates, equals to

$$ \dim S_{n|r} - \dim SL(n) = \dfrac{(n + r - 1)!}{r! (n-1)!} - n^2 + 1 $$
\smallskip\\
Several examples of these numbers are shown at Fig. 3.

\begin{figure}[h]
\begin{center}
$
\begin{array}{c|ccccccc}
r \backslash n & & 2 & 3 & 4 & 5 & 6 & 7 \\
\\
\hline\\
2 & & 1 & 1 & 1 & 1 & 1 & 1 \\
\\
3 & & 1 & 2 & 5 & 11 & 21 & 36 \\
\\
4 & & 2 & 7 & 20 & 46 & 91 & 162 \\
\\
5 & & 3 & 13 & 41 & 102 & 217 & 414 \\
\\
6 & & 4 & 20 & 69 & 186 & 427 & 876 \\
\\
\end{array}
$
\caption{The number of functionally independent $SL(n)$ invariants $I_{k}$ of a form of degree $r$ in $n$ variables. }
\end{center}
\end{figure}

\paragraph{The Gaussian case $J_{n|2}$.}Note, however, that the case $r = 2$ (quadratic forms) is exceptional, because the dimension $\dim S_{n|2} = n(n+1)/2$ is less than the dimension $\dim SL(n) = n^2 - 1$. For this reason, the above dimension counting does not work in this (and only in this) case. Actually, as we know, the space of $SL(n)$ orbits on quadratic forms is one-dimensional, and the only invariant -- the single coordinate on the space of orbits -- is determinant $\det S$. Thus, determinant is the only variable $J_{n|2}$ can depend on:

$$J_{n|2}\big( S \big) = F\big( \det S \big) $$
\smallskip\\
The homogeneity condition states, that $F\big( \lambda x \big) = F\big( x \big)/\sqrt{\lambda}$ and has a single solution $F(x) = 1/\sqrt{x}$.
In this way we reproduce the well-known Gaussian integral

\begin{equation}
\addtolength{\fboxsep}{5pt}
\boxed{
\begin{gathered}
\ \ \ J_{n|2}\big( S \big) = \dfrac{1}{\sqrt{\det S}} \ \ \
\end{gathered}
}\label{Gaussian}
\end{equation}
\smallskip\\
We emphasize, that simplicity of this answer is due to simplicity of the space of orbits, i.e, due to the fact there is a single invariant in this case. For higher $r$, there are many invariants (as many as shown at Fig. 3) and $J_{n|r}$ is a non-trivial function of all of them. The problem is that it becomes impossible to find $J_{n|r}$ from the homogeneity condition alone. The solution of this problem is provided by Ward identities: in addition to the homogeneity condition, integral discriminant satisfies relations (\ref{Ward}) and this allows to find $F\big\{ I_k \big\}$.

\paragraph{Diagram technique for $SL(n)$ invariants.} To do actual computations with $SL(n)$ invariants, we will adopt a convenient diagram technique, which is described in \cite{nolal}. According to \cite{nolal}, it is possible to represent tensors with $k$ indices as $k$-valent vertices, with two types of indices -- covariant and contravariant -- represented by two types of lines. Contraction of indices is naturally represented as connection of these lines. To simplify the diagrams, we use solid lines for both covariant and contravariant indices. In this paper we use the following four \emph{elementary building blocks} for diagrams: covariant tensor $S_{i_1 \ldots i_r}$, represented by a black $r$-valent vertex
\begin{center}
\includegraphics[totalheight=90pt]{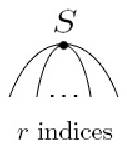}
\end{center}
contravariant tensor $\dfrac{\partial}{\partial S_{i_1 \ldots i_r}}$, represented by a black $r$-valent vertex
\begin{center}
\includegraphics[totalheight=90pt]{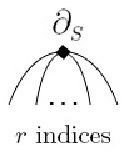}
\end{center}
completely antisymmetric contravariant tensor $\epsilon^{i_1 \ldots i_n}$, represented by a white $n$-valent vertex
\begin{center}
\includegraphics[totalheight=90pt]{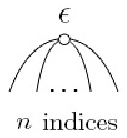}
\end{center}
and completely antisymmetric covariant tensor $\epsilon^{*}_{i_1 \ldots i_n}$, represented by a white $n$-valent vertex
\begin{center}
\includegraphics[totalheight=90pt]{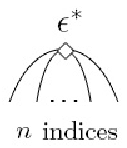}
\end{center}
Tensors $\epsilon^{i_1 \ldots i_n}$ and $\epsilon^{*}_{i_1 \ldots i_n}$ are completely antisymmetric with respect to permutations of their indices, and, therefore, remain invariant under $SL(n)$ transformations:

$$\epsilon^{i_1 \ldots i_n} U^{j_1}_{i_1} \ldots U^{j_n}_{i_n} = \det U \cdot \epsilon^{j_1 \ldots j_n} = \epsilon^{j_1 \ldots j_n}$$

$$\epsilon^{*}_{i_1 \ldots i_n} U^{i_1}_{j_1} \ldots U^{i_n}_{j_n} = \det U \cdot \epsilon^{*}_{j_1 \ldots j_n}  = \epsilon^{*}_{j_1 \ldots j_n}$$
\smallskip\\
where $\det U = 1$, because $U \in SL(n)$. For this reason, any diagram without free (uncontracted) indices, made of contravariant $\epsilon$-vertices and covariant $S$-vertices, is automatically $SL(n)$-invariant function of $S$. Say, determinant of $n \times n$ matrix $S_{ij}$ can be represented as a diagram at Fig. 4, with $n$ $S$-vertices and two $\epsilon$-vertices. The other two types of vertices ($\partial/\partial S$ and $\epsilon^{*}$) will be used to construct $SL(n)$-invariant differential operators.

\begin{figure}[h]
\begin{center}
\includegraphics[totalheight=130pt]{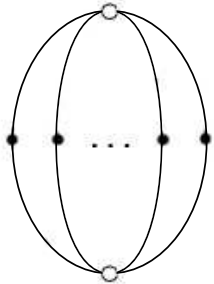}
\caption{Determinant of matrix $S_{ij}$, represented as a diagram of tensor contraction. Black 2-valent vertices represent tensor $S$, white n-valent vertices represent tensor $\epsilon$.}
\end{center}
\end{figure}

Diagrams provide a convenient way to obtain $SL(n)$ invariants. In fact, any $SL(n)$-invariant can be obtained in this way, but, unfortunately, not \emph{uniquely}: absolutely different-looking diagrams can represent one and the same invariant. Trying to resolve this ambiguity, one typically faces the complicated problems of classification of diagrams and finding relations between diagrams. These problems will not be adressed (and even touched) in present paper. Our goal is different: to find an explicit answer for $J_{n|r}$ in several non-Gaussian cases, using diagrams as a convenient tool. We now turn to accomplishing this goal.

\subsection{The case of $J_{2|3}$}

The simplest non-trivial (i.e. non-Gaussian) example is a $3$-form in $2$ variables, which can be written as

$$ S(x,y) = S_{111} x^3 + 3 S_{112} x^2 y + 3 S_{122} x y^2 + S_{222} y^2 $$
\smallskip\\
By dimension counting, there is only one elementary invariant $I_{4}$ in this case, given by a diagram at Fig. 5. \begin{figure}[t]
\begin{center}
\includegraphics[totalheight=160pt,clip]{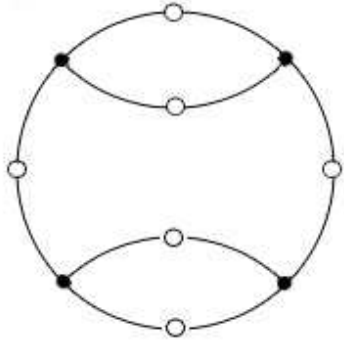}
\caption{The degree 4 invariant $I_4$ of a 3-form in 2 variables, represented as a diagram of tensor contraction. Black 3-valent vertices represent tensor $S$, white 2-valent vertices represent tensor $\epsilon$.}
\end{center}
\end{figure} The subscript "4" stands for the degree of this invariant. In this paper we find it convenient to denote the elementary invariants of degree $k$ as $I_{k}$. It is straightforward to write the algebraic expression for the diagram:

$$ I_{4} = S_{i_1 i_2 i_3} S_{j_1 j_2 j_3} S_{k_1 k_2 k_3} S_{l_1 l_2 l_3} \epsilon^{i_1 j_1} \epsilon^{i_2 j_2} \epsilon^{k_1 l_1} \epsilon^{k_2 l_2} \epsilon^{i_3 k_3} \epsilon^{j_3 l_3} $$
\smallskip\\
Evaluating this sum, one gets the following explicit formula for $I_4$

$$ I_{4} = 2 S_{111}^2 S_{222}^2-12 S_{111} S_{112} S_{122} S_{222}+8 S_{111} S_{122}^3+8 S_{112}^3 S_{222}-6 S_{112}^2 S_{122}^2$$
\smallskip\\
which is nothing but the algebraic discriminant $ D_{2|3} $ of $S$:
\begin{align}
D_{2|3} = I_4
\label{D23}
\end{align}
Since there is only one elementary invariant, the integral discriminant $J_{2|3}$ must be a function of $D_{2|3}$:
$$J_{2|3}\big( S \big) = F\big( I_{4} \big) $$
Thus, this case is similar to the Gaussian case. The homogeneity condition states, that $F\big( \lambda x \big) = F\big( x \big)/\sqrt[6]{\lambda}$ and has a single solution $F(x) = 1/\sqrt[6]{x}$. In this way we reproduce the formula from the Introduction:

\begin{equation}
\addtolength{\fboxsep}{5pt}
\boxed{
\begin{gathered}
\ \ \ J_{2|3}\big( S \big) = I_{4}^{-1/6} \ \ \
\end{gathered}
}\label{23}
\end{equation}
\smallskip\\
In this case, one does not need to use the Ward identities (\ref{Ward}): it is enough to use the homogeneity condition.

\begin{figure}[t]
\begin{center}
\includegraphics[totalheight=175pt,clip]{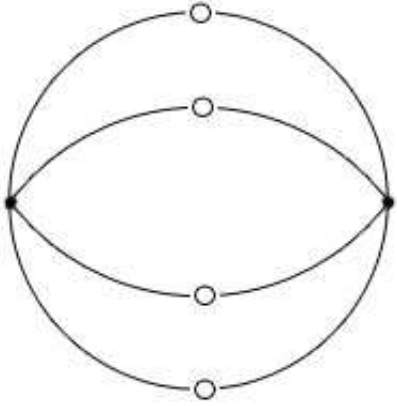}
\caption{The degree 2 invariant $I_2$ of a 4-form in 2 variables, represented as a diagram of tensor contraction. Black 4-valent vertices represent tensor $S$, white 2-valent vertices represent tensor $\epsilon$.}
\end{center}
\end{figure}

\begin{figure}[t]
\begin{center}
\includegraphics[totalheight=175pt,clip]{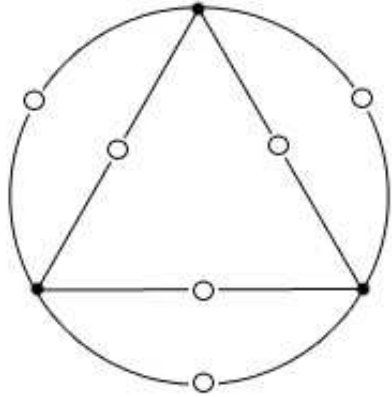}
\caption{The degree 3 invariant $I_3$ of a 4-form in 2 variables, represented as a diagram of tensor contraction. Black 4-valent vertices represent tensor $S$, white 2-valent vertices represent tensor $\epsilon$.}
\end{center}
\end{figure}

\subsection{The case of $J_{2|4}$}

The next-to-simplest example is a $4$-form in $2$ variables, which can be written as

$$ S(x,y) = S_{1111} x^4 + 4 S_{1112} x^3 y + 6 S_{1122} x^2 y^2 + 4 S_{1222} x y^3 + S_{2222} y^4 $$
\paragraph{Invariants.}By dimension counting, there are two elementary invariants in this case. They have relatively low degrees $2$ and $3$, denoted as $I_{2}$ and $I_3$ and given by diagrams at Fig. 6 and Fig. 7, respectively. Looking at the diagrams, it is straightforward to write algebraic expressions for $I_2, I_3$:

$$ I_{2} = S_{i_1 i_2 i_3 i_4} S_{j_1 j_2 j_3 j_4} \epsilon^{i_1 j_1} \epsilon^{i_2 j_2} \epsilon^{i_3 j_3} \epsilon^{i_4 j_4}$$

$$ I_{3} = S_{i_1 i_2 i_3 i_4} S_{j_1 j_2 j_3 j_4} S_{k_1 k_2 k_3 k_4} \epsilon^{i_1 j_1} \epsilon^{i_2 j_2} \epsilon^{i_3 k_1} \epsilon^{i_4 k_2} \epsilon^{j_3 k_3} \epsilon^{j_4 k_4}$$
\smallskip\\
Evaluating these sums, one gets the following explicit formulas for $I_2, I_3$:

$$ I_{2} = 2 S_{1111} S_{2222} - 8 S_{1112} S_{1222} + 6 S_{1122}^2$$

$$ I_{3} = 6 S_{1111} S_{1122} S_{2222}-6 S_{1111} S_{1222}^2-6 S_{1112}^2 S_{2222}+12 S_{1112} S_{1122} S_{1222}-6 S_{1122}^3$$
\smallskip\\
The algebraic discriminant $D_{2|4}$, just like any other $SL(2)$-invariant function of $S$, is a function of $I_2,I_3$:

\begin{align}
D_{2|4} = & \ I_2^3 - 6 I_{3}^2 \ = 8 S_{1111}^3 S_{2222}^3-96 S_{1111}^2 S_{1112} S_{1222} S_{2222}^2-144 S_{1111}^2 S_{1122}^2 S_{2222}^2+ 432 S_{1111}^2 S_{1122} S_{1222}^2 S_{2222}\nonumber \\ & \nonumber -216 S_{1111}^2 S_{1222}^4+432 S_{1111} S_{1112}^2 S_{1122} S_{2222}^2-48 S_{1111} S_{1112}^2 S_{1222}^2 S_{2222}-1440 S_{1111} S_{1112} S_{1122}^2 S_{1222} S_{2222} \\ & \nonumber + 648 S_{1111} S_{1122}^4 S_{2222} + 864 S_{1111} S_{1112} S_{1122} S_{1222}^3 +864 S_{1112}^3 S_{1122} S_{1222} S_{2222} +288 S_{1112}^2 S_{1122}^2 S_{1222}^2 \\ & -432 S_{1111} S_{1122}^3 S_{1222}^2-216 S_{1112}^4 S_{2222}^2-512 S_{1112}^3 S_{1222}^3-432 S_{1112}^2 S_{1122}^3 S_{2222}
\label{D24}
\end{align}
\paragraph{Integral discriminant.}Similarly, the integral discriminant is a function of $I_{2}, I_{3}$:

$$J_{2|4} = F\big( I_2, I_3 \big)$$
\smallskip\\
where the function $F$ is to be determined. The homogeneity condition does not allow to find $F$ unambigously, it constrains only the overall scaling factor, but not the non-trivial dependence on the ratio of invariants:

\begin{align}
F\big( I_2, I_3 \big) = I_{2}^{-1/4} \ G\left( \dfrac{I_3^2}{I_2^3} \right)
\label{Ansatz24}
\end{align}

\paragraph{Ward identities.}To find the function $G(z)$ in this ansatz, we need to use the Ward identities (\ref{Ward}). Applied to the present case of $n = 2, r = 4$, the system (\ref{Ward}) takes the form

\[
\begin{array}{ccc}
\left( \dfrac{\partial^2}{\partial s_{40} \partial s_{22}} - \dfrac{\partial^2}{\partial s_{31} \partial s_{31} } \right) J_{2|4} = 0 \\
\\
\left( \dfrac{\partial^2}{\partial s_{40} \partial s_{13}} - \dfrac{\partial^2}{\partial s_{31} \partial s_{22} } \right) J_{2|4} = 0 \\
\\
\left( \dfrac{\partial^2}{\partial s_{40} \partial s_{04}} - \dfrac{\partial^2}{\partial s_{31} \partial s_{13} } \right) J_{2|4} = 0 \\
\\
\left( \dfrac{\partial^2}{\partial s_{22} \partial s_{22}} - \dfrac{\partial^2}{\partial s_{31} \partial s_{13} } \right) J_{2|4} = 0 \\
\\
\end{array}
\]
where $s$-parameters and $S$-parameters are related by

$$ s_{40} = S_{1111}, \ s_{31} = 4 S_{1112}, \ s_{22} = 6 S_{1122}, \ s_{13} = 4 S_{1222}, \ s_{04} = S_{2222} $$
\smallskip\\
Particular equations in this system are, of course, not $SL(2)$ invariant. This is not convenient, since we are interested in $SL(2)$-invariant solutions. Let us transform Ward identities into invariant form, using (\ref{Ansatz24}).

\paragraph{Invariant form of Ward identities.}Substituting the invariant anzatz (\ref{Ansatz24}) into the system, we obtain
\[
\begin{array}{ccc}
\\
\left( \dfrac{\partial^2}{\partial s_{40} \partial s_{22}} - \dfrac{\partial^2}{\partial s_{31} \partial s_{31} } \right) J_{2|4} = \dfrac{S_{1122} S_{2222} - S_{1222}^2}{4 I_2^{9/4}} \cdot \left( ( 144 z^2 - 24 z ) \dfrac{\partial^2 G(z)}{\partial z^2} + \left( 216 z - 12 \right) \dfrac{\partial G(z)}{\partial z} + 5 G(z) \right) \\
\\
\left( \dfrac{\partial^2}{\partial s_{40} \partial s_{13}} - \dfrac{\partial^2}{\partial s_{31} \partial s_{22} } \right) J_{2|4} = \dfrac{S_{1222}S_{1122} - S_{1112} S_{2222} }{4 I_2^{9/4}} \cdot \left( ( 144 z^2 - 24 z ) \dfrac{\partial^2 G(z)}{\partial z^2} + \left( 216 z - 12 \right) \dfrac{\partial G(z)}{\partial z} + 5 G(z) \right) \\
\\
\left( \dfrac{\partial^2}{\partial s_{40} \partial s_{04}} - \dfrac{\partial^2}{\partial s_{31} \partial s_{13} } \right) J_{2|4} = \dfrac{3 S_{1112} S_{1222} - 3 S_{1122}^2}{4 I_2^{9/4}} \cdot \left( ( 144 z^2 - 24 z ) \dfrac{\partial^2 G(z)}{\partial z^2} + \left( 216 z - 12 \right) \dfrac{\partial G(z)}{\partial z} + 5 G(z) \right) \\
\\
\left( \dfrac{\partial^2}{\partial s_{22} \partial s_{22}} - \dfrac{\partial^2}{\partial s_{31} \partial s_{13} } \right) J_{2|4} = \dfrac{S_{1111} S_{2222} - S_{1112} S_{1222}}{12 I_2^{9/4}} \cdot \left( ( 144 z^2 - 24 z ) \dfrac{\partial^2 G(z)}{\partial z^2} + \left( 216 z - 12 \right) \dfrac{\partial G(z)}{\partial z} + 5 G(z) \right) \\
\\
\end{array}
\]
where $z = I_3^2 / I_2^3$.  These equations contain a common $SL(2)$ invariant factor. We conclude, that all the four expressions vanish, if and only if $G(z)$ satisfies the differential equation

\begin{align}
( 144 z^2 - 24 z ) \dfrac{\partial^2 G(z)}{\partial z^2} + \left( 216 z - 12 \right) \dfrac{\partial G(z)}{\partial z} + 5 G(z) = 0
\label{24hypergeom}
\end{align}
\smallskip\\
which is nothing but Gauss hypergeometric equation

$$
t(1 - t) \dfrac{\partial^2 G(t)}{\partial t^2} + \left( c - (a + b + 1) t \right) \dfrac{\partial G(t)}{\partial t} - ab G(t) = 0
$$
\smallskip\\
with $a = 1/12$, $b = 5/12$, $c = 1/2$ and $t = 6z$. This is the invariant form of $n = 2, r = 4$ Ward identities.

\paragraph{Solution.} In terms of the Gauss hypergeometric function

$${}_{2}F_{1} \big( \left[ a, b \right], \left[ c \right], t \big) = \sum\limits_{k = 0}^{\infty} \dfrac{\Gamma(a + k)}{\Gamma(a)} \dfrac{\Gamma(b + k)}{\Gamma(b)} \dfrac{\Gamma(c)}{\Gamma(c + k)} \dfrac{t^k}{k!} = 1 + \dfrac{ab}{c} t + \dfrac{a(a+1)b(b+1)}{c(c+1)} \dfrac{t^2}{2} + \ldots $$
\smallskip\\
the general solution of Gauss hypergeometric equation is given by

$$
G(t) = c_1 \cdot {}_{2}F_{1} \big( \left[ a, b \right], \left[ c \right], t \big) + c_2 \cdot t^{1-c} \ {}_{2}F_{1} \big( \left[ a + 1 - c, b + 1 - c\right], \left[ 2 - c \right], t \big)
$$
\smallskip\\
Consequently, the integral discriminant equals

\begin{equation}
\addtolength{\fboxsep}{5pt}
\boxed{
\begin{gathered}
\ \ \ J_{2|4}\big(S\big) = c_1 \cdot I_{2}^{-1/4} {}_{2}F_{1} \left( \left[ \dfrac{1}{12}, \dfrac{5}{12} \right], \left[ \dfrac{1}{2} \right], \dfrac{6I_{3}^2}{I_{2}^3} \right)
+ c_2 \cdot I_{3} I_{2}^{-7/4}  {}_{2}F_{1} \left( \left[ \dfrac{7}{12}, \dfrac{11}{12} \right], \left[ \dfrac{3}{2} \right], \dfrac{6I_{3}^2}{I_{2}^3} \right) \ \ \
\end{gathered}
}\label{24}
\end{equation}
\smallskip\\
where $c_{1,2}$ are the two constants, parametrising the general solution of Ward identities. Particular solutions

$$
J^{(1)}_{2|4}\big(S\big) = I_{2}^{-1/4} {}_{2}F_{1} \left( \left[ \dfrac{1}{12}, \dfrac{5}{12} \right], \left[ \dfrac{1}{2} \right], \dfrac{6I_{3}^2}{I_{2}^3} \right) \ \ \ \mbox{ and } \ \ \ J^{(2)}_{2|4}\big(S\big) = I_{3} I_{2}^{-7/4} {}_{2}F_{1} \left( \left[ \dfrac{7}{12}, \dfrac{11}{12} \right], \left[ \dfrac{3}{2} \right], \dfrac{6I_{3}^2}{I_{2}^3} \right)
$$
\smallskip\\
are associated with different integration contours and can be called the first and the second branches of $J_{2|4}$.

\paragraph{Singularities.}Notice, that the point $t = 1$ corresponds to

$$ \dfrac{6I_{3}^2}{I_{2}^3} = 1$$
\smallskip\\
which is just the discriminant's vanishing locus $ I_{2}^3 - 6I_{3}^2 = D_{2|4} = 0$. This is interesting, because the point $t = 1$ is a singular point of the hypergeometric function ${}_{2}F_{1}$. However, there are two other suspicious points: hypergeometric function ${}_{2}F_{1}$ can have singularities at $t = 0, 1$ and $\infty$. Let us study asymptotics of integral discriminants at these points, using the formulas

\begin{align}
{}_{2}F_{1} \left( \left[ a, b \right], \left[ c \right], t \right) = 1 + O(t), \ \ \ \ \mbox{ when } t \rightarrow 0
\label{HG1}
\end{align}

\begin{align}
{}_{2}F_{1} \left( \left[ a, b \right], \left[ a + b \right], 1 - t \right) = - \dfrac{\Gamma(a + b)}{\Gamma(a)\Gamma(b)} \log t + O\big(t^0\big), \ \ \ \ \mbox{ when } t \rightarrow 0
\label{HG2}
\end{align}

\begin{align}
\nonumber {}_{2}F_{1} \left( [a,b],[c], t \right) \ = \ & t^{-a} \dfrac{\Gamma( b - a )\Gamma( c )}{\Gamma( b ) \Gamma( c - a )} \cdot {}_{2}F_{1} \left( [a,a - c + 1],[a - b + 1], 1/t \right) + \emph{} \\ & \nonumber \\ & t^{-b} \dfrac{\Gamma( a - b )\Gamma( c )}{\Gamma( a ) \Gamma( c - b )} \cdot {}_{2}F_{1} \left( [b,b - c + 1],[b - a + 1], 1/t \right)
\label{HG4}
\end{align}
\smallskip\\
which can be found in any reference book of hypergeometric functions (see, e.g. \cite{hypergeom}). Using (\ref{HG1}), we find

$$
J^{(1)}_{2|4}\big(S\big) \sim I_{2}^{-1/4} + O\big(t\big), \ \ \ \ \mbox{ when } t \rightarrow 0
$$

$$
J^{(2)}_{2|4}\big(S\big) \sim I_{3} I_{2}^{-7/4} + O\big(t\big), \ \ \ \ \mbox{ when } t \rightarrow 0
$$
\smallskip\\
Using (\ref{HG2}), we find

$$
J^{(1)}_{2|4}\big(S\big) \sim \dfrac{- \Gamma(1/2)}{\Gamma(1/12)\Gamma(5/12)} I_{2}^{-1/4} \log \big( 1 - \dfrac{6I_{3}^2}{I_2^3} \big) + O\big(t^0\big), \ \ \ \ \mbox{ when } t \rightarrow 1
$$

$$
J^{(2)}_{2|4}\big(S\big) \sim \dfrac{-\Gamma(3/2)}{\Gamma(7/12)\Gamma(11/12)} I_{3} I_{2}^{-7/4} \log \big( 1 - \dfrac{6I_{3}^2}{I_2^3} \big) + O\big(t^0\big), \ \ \ \ \mbox{ when } t \rightarrow 1
$$
\smallskip\\
Using (\ref{HG1}) and (\ref{HG4}), we find

$$
J^{(1)}_{2|4}\big(S\big) \sim \dfrac{\Gamma(1/2) \Gamma(1/3)}{\Gamma(5/12)^2} (-6)^{-1/12} I_{3}^{-1/6} + O\big(t^{-5/12}\big), \ \ \ \ \mbox{ when } t \rightarrow \infty
$$

$$
J^{(2)}_{2|4}\big(S\big) \sim \dfrac{\Gamma(1/3) \Gamma(3/2)}{\Gamma(11/12)^2} (-6)^{-7/12} I_{3}^{-1/6} + O\big(t^{-11/12}\big), \ \ \ \ \mbox{ when } t \rightarrow \infty
$$
\smallskip\\
We can see from these results, that singularities of the integral discriminant $J_{2|4}$ at finite values of $I_{2}, I_{3}$ are completely controlled by the algebraic discriminant $D_{2|4}$. The singularity at discriminant locus is logarithmic. Other singularities, not related to the algebraic discriminant, are situated at infinite values of invariants: $I_{2} \rightarrow \infty$ ($t \rightarrow 0$) and $I_{3} \rightarrow \infty$ ($t \rightarrow \infty$).

\paragraph{Hypergeometric integral.}Function ${}_{2}F_{1}$ has an integral representation

\begin{align}
 {}_{2}F_{1} \big( \left[ a, b \right], \left[ c \right], t \big) = \dfrac{\Gamma(c)}{\Gamma(b)\Gamma(c-b)} \cdot \int\limits_{0}^{1} ds \dfrac{s^{b-1}(1-s)^{c-b-1}}{(1-st)^{a}}
\label{HyperInt}
\end{align}
\smallskip\\
which is a direct consequence of a simpler identity

\begin{align}
 \dfrac{ \Gamma( a ) \Gamma( b )}{\Gamma( a + b )} = \int\limits_{0}^{1} ds \ s^{a - 1} (1 - s)^{b - 1}
\label{BFunction}
\end{align}
\smallskip\\
For the integral discriminant, we obtain

\begin{equation}
\addtolength{\fboxsep}{5pt}
\boxed{
\begin{gathered}
J_{2|4}\big(S\big) = c_1 \cdot \int\limits_{0}^{1} ds \ \dfrac{ s^{-7/12}(1-s)^{-11/12}}{(I_2^3 - 6 s I_3^2)^{1/12}} + c_2 I_3 \cdot \int\limits_{0}^{1} ds \ \dfrac{s^{-1/12}(1-s)^{-5/12}}{(I_2^3 - 6 s I_3^2)^{7/12}}
\end{gathered}
}\label{integ24}
\end{equation}
\smallskip\\
Hypergeometric functions with different values of parameters $a,b$ and $c$ are related by various "modular" transformations of the variable $t$ and the integration variable $s$, which leave the boundary region (two points 0 and 1) intact. Such transformations were first found and studied by Euler, therefore they are known as Euler hypergeometric transformations. Say, transformation $ t \mapsto t, \ s \mapsto 1 - s $ gives rise to a relation

\begin{align}
{}_{2}F_{1} \left( \left[ a, b \right], \left[ c \right], t \right) = (1 - t)^{-a} \cdot {}_{2}F_{1} \left( \left[ a, c - b \right], \left[ c \right], \dfrac{t}{t - 1} \right)
\label{HG3}
\end{align}
\smallskip\\
After this transformation, $J_{2|4}$ takes form

$$
J^{(1)}_{2|4}\big(S\big) = \left( D_{2|4} \right)^{-1/12} {}_{2}F_{1} \left( \left[ \dfrac{1}{12}, \dfrac{1}{12} \right], \left[ \dfrac{1}{2} \right], - \dfrac{6I_{3}^2}{D_{2|4}} \right)
$$

$$
J^{(2)}_{2|4}\big(S\big) = I_{3} \left( D_{2|4}\right)^{-7/12} {}_{2}F_{1} \left( \left[ \dfrac{7}{12}, \dfrac{7}{12} \right], \left[ \dfrac{3}{2} \right], - \dfrac{6I_{3}^2}{D_{2|4}} \right)
$$
\smallskip\\
In this way the hypergeometric integral (\ref{integ24}) allows to recast $J_{2|4}$ in various forms and establish relations between them. Therefore, (\ref{integ24}) is a useful and important representation. We emphasise, that the relation of integral $\int e^{-S(x,y)} dx dy$ to hypergeometric integrals (\ref{integ24}) is not \emph{a priori} expected: these integrals look very different, even from the point of view of variables they depend on.
\paragraph{Vertical symmetry.}To finish this section, let us check the vertical symmetry between $J_{2|2}$ and $J_{2|4}$:

$$ J_{2|4} \Big( (ax^2 + bxy + cy^2)^2 \Big) \sim J_{2|2} \Big( ax^2 + bxy + cy^2 \Big) $$
\smallskip\\
Setting $S(x,y) = (ax^2 + bxy + cy^2)^2$ we obtain the coefficients

\[
\begin{array}{ccc}
S_{1111} = a^2, \ \ S_{1112} = \dfrac{1}{2} ab, \ \ S_{1122} = \dfrac{1}{3} ac + \dfrac{1}{6} b^2, \ \ S_{1222} = \dfrac{1}{2} bc, \ \ S_{2222} = c^2\\
\end{array}
\]
\smallskip\\
and the invariants
\[
\begin{array}{ccc}
I_2 = \dfrac{1}{6} (b^2 - 4ac)^2, \ \ I_3 = \dfrac{1}{36} (b^2 - 4ac)^3 \\
\end{array}
\]
\smallskip\\
Substituting them into (\ref{24}), we find

\[
\begin{array}{ccc}
J^{(1)}_{2|4}\big(S\big) = \dfrac{A^{(1)}}{\sqrt{b^2 - 4ac}} \\
\\
J^{(2)}_{2|4}\big(S\big) = \dfrac{A^{(2)}}{\sqrt{b^2 - 4ac}} \\
\end{array}
\]
\smallskip\\
where the constants of proportionality

\[
\begin{array}{ccc}
A^{(1)} = 6^{-1/4} \cdot {}_{2}F_{1} \left( \left[ \dfrac{1}{12}, \dfrac{5}{12} \right], \left[ \dfrac{1}{2} \right], 1 \right) \sim -6^{-1/4} \dfrac{\Gamma(1/2)}{\Gamma(1/12)\Gamma(5/12)} \log 0 \sim \infty \\
\\
A^{(2)} = 6^{+1/4} \cdot {}_{2}F_{1} \left( \left[ \dfrac{7}{12}, \dfrac{11}{12} \right], \left[ \dfrac{3}{2} \right], 1 \right) \sim -6^{+1/4} \dfrac{\Gamma(3/2)}{\Gamma(7/12)\Gamma(11/12)} \log 0 \sim \infty \\
\end{array}
\]
\smallskip\\
are independent of $a,b,c$ -- as prescribed by the vertical symmetry -- but infinite. Only their linear combination

$$\left( 6^{+1/4} \dfrac{\Gamma(3/2)}{\Gamma(7/12)\Gamma(11/12)} \right) A^{(1)} - \left( 6^{-1/4} \dfrac{\Gamma(1/2)}{\Gamma(1/12)\Gamma(5/12)} \right) A^{(2)} = \mbox{ finite number } = -\dfrac{1}{2}$$
\smallskip\\
is finite -- logarithmic divergencies cancel out. Actually, it is expectable that only one linear combination of solutions stays finite. This is because not every contour, admissible for $J_{2|4}$, is admissible for $J_{2|2}$. Admissible contours for $J_{2|4}$ approach infinity from 4 different directions, where the 4-form $(ax^2 + bxy + cy^2)^2$ takes real positive values. Only two of these directions are appropriate for $J_{2|2}$ -- those, where the 2-form $ax^2 + bxy + cy^2$ takes real positive values. One linear combination of branches corresponds to admissible contours, while the orthogonal linear combination corresponds to other contours. That is why both branches are singular at $t \rightarrow 1$, but certain linear combination of branches is regular at $t = 1$. Note, that nothing similar happens at non-discriminantal singularities: at $t \rightarrow 0$ and $t \rightarrow \infty$ no linear combination of branches stays regular.

\begin{figure}[h]
\begin{center}
\includegraphics[totalheight=220pt,clip]{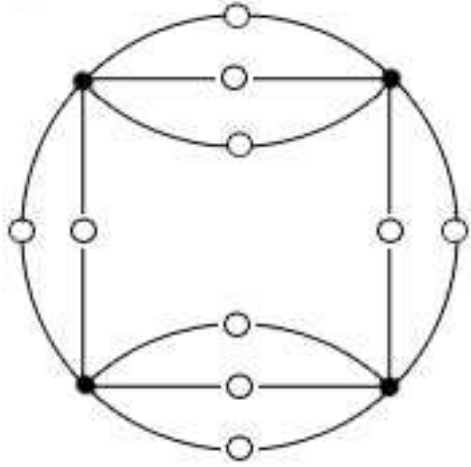}
\caption{The degree 4 invariant $I_4$ of a 5-form in 2 variables, represented as a diagram of tensor contraction. Black 5-valent vertices represent tensor $S$, white 2-valent vertices represent tensor $\epsilon$.}
\end{center}
\end{figure}

\subsection{The case of $J_{2|5}$}

Our third example is a $5$-form in $2$ variables, which can be written as

$$ S(x,y) = S_{11111} x^5 + 5 S_{11112} x^4 y + 10 S_{11122} x^3 y^2 + 10 S_{11222} x^2 y^3 + 5 S_{12222} x y^4 + S_{22222} y^5 $$
\paragraph{Invariants.} By dimension counting, there are three elementary invariants in this case. They have degrees $4$, $8$ and $12$, denoted as $I_{4}$, $I_{8}$ and $I_{12}$ and given by diagrams at Fig. 8, Fig. 9 and Fig. 10. Looking at the diagrams, it is straightforward to write an expression for $I_{4}$

$$ I_{4} = S_{i_1 i_2 i_3 i_4 i_5} S_{j_1 j_2 j_3 j_4 j_5} S_{k_1 k_2 k_3 k_4 k_5} S_{l_1 l_2 l_3 l_4 l_5} \epsilon^{i_1 j_1} \epsilon^{i_2 j_2} \epsilon^{i_3 j_3} \epsilon^{i_4 k_4} \epsilon^{i_5 k_5} \epsilon^{j_4 l_4} \epsilon^{j_5 l_5} \epsilon^{k_1 l_1} \epsilon^{k_2 l_2} \epsilon^{k_3 l_3} $$
\smallskip\\
and equally straightforward to write expressions for $I_{8}, I_{12}$. Evaluating the contraction, one gets a formula

\begin{align*}
I_{4} \ = \ & 2 S_{11111}^2 S_{22222}^2-20 S_{11111} S_{11112} S_{12222} S_{22222}+8 S_{11111} S_{11122} S_{11222} S_{22222}+ \emph{} \\ & 32 S_{11111} S_{11122} S_{12222}^2-24 S_{11111} S_{11222}^2 S_{12222}+32 S_{11112}^2 S_{11222} S_{22222}+18 S_{11112}^2 S_{12222}^2-\emph{} \\ &24 S_{11112} S_{11122}^2 S_{22222}-152 S_{11112} S_{11122} S_{11222} S_{12222}+96 S_{11112} S_{11222}^3+96 S_{11122}^3 S_{12222}-64 S_{11122}^2 S_{11222}^2
\end{align*}
\smallskip\\
and similar formulas for $I_{8}, I_{12}$ -- they are quite lengthy and we do not present them here. The algebraic discriminant $D_{2|5}$, just like any other $SL(2)$-invariant function of $S$, is a function of $I_4,I_8,I_{12}$:

\begin{align}
D_{2|5} = I_4^2 - 64 I_{8}
\label{D25}
\end{align}
\paragraph{Integral discriminant.}Similarly, the integral discriminant is a function of $I_4,I_8,I_{12}$:

$$J_{2|5} = F\big( I_4,I_8,I_{12} \big)$$
\smallskip\\
where the function $F$ is to be determined. The homogeneity condition does not allow to find $F$ unambigously, it constrains only the overall scaling factor, but not the non-trivial dependence on the ratios of invariants:

\begin{figure}[t]
\begin{center}
\includegraphics[totalheight=200pt,clip]{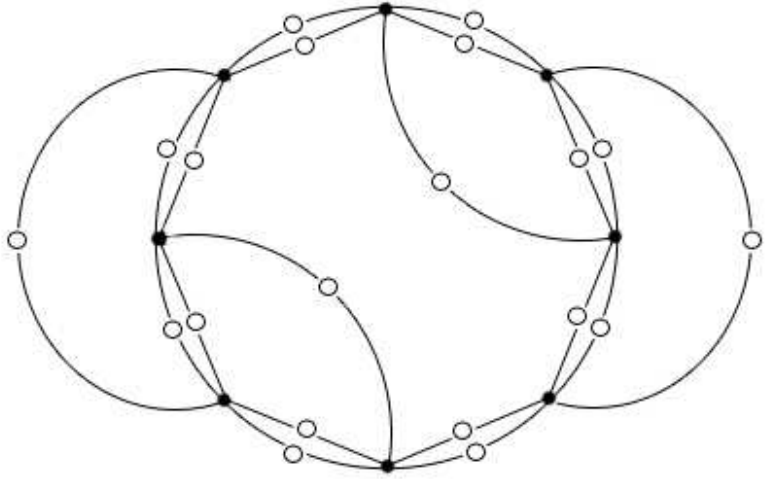}
\caption{The degree 8 invariant $I_8$ of a 5-form in 2 variables, represented as a diagram of tensor contraction. Black 5-valent vertices represent tensor $S$, white 2-valent vertices represent tensor $\epsilon$.}
\end{center}
\end{figure}

\begin{align}
F\big( I_4,I_8,I_{12} \big) = I_{4}^{-1/10} \ G\left( \dfrac{I_8}{I_4^2}, \dfrac{I_{12}}{I_4^3} \right)
\label{Ansatz25}
\end{align}
\smallskip\\
To find the function $G(u,v)$ in this ansatz, we need to use the Ward identities (\ref{Ward}).

\paragraph{Ward identities.} Applied to the present case of $n = 2, r = 5$, the system (\ref{Ward}) takes the form
\[
\begin{array}{ccc}
\\
\left( \dfrac{\partial^2}{\partial s_{50} \partial s_{32}} - \dfrac{\partial^2}{\partial s_{41} \partial s_{41} } \right) J_{2|5} = 0 \\
\\
\left( \dfrac{\partial^2}{\partial s_{50} \partial s_{23}} - \dfrac{\partial^2}{\partial s_{41} \partial s_{32} } \right) J_{2|5} = 0 \\
\\
\left( \dfrac{\partial^2}{\partial s_{50} \partial s_{14}} - \dfrac{\partial^2}{\partial s_{41} \partial s_{23} } \right) J_{2|5} = 0 \\
\\
\end{array}
\begin{array}{ccc}
\\
\left( \dfrac{\partial^2}{\partial s_{32} \partial s_{32}} - \dfrac{\partial^2}{\partial s_{41} \partial s_{23} } \right) J_{2|5} = 0 \\
\\
\left( \dfrac{\partial^2}{\partial s_{50} \partial s_{05}} - \dfrac{\partial^2}{\partial s_{32} \partial s_{23} } \right) J_{2|5} = 0 \\
\\
\left( \dfrac{\partial^2}{\partial s_{41} \partial s_{14}} - \dfrac{\partial^2}{\partial s_{32} \partial s_{23} } \right) J_{2|5} = 0 \\
\\
\end{array}
\]
where $s$-parameters and $S$-parameters are related by

$$ s_{50} = S_{11111}, \ s_{41} = 5 S_{11112}, \ s_{32} = 10 S_{11122}, \ s_{23} = 10 S_{11222}, \ s_{14} = 5 S_{12222}, \ s_{05} = S_{22222} $$
\smallskip\\
Particular equations in this system are, of course, not $SL(2)$ invariant.
\begin{figure}[t]
\begin{center}
\includegraphics[totalheight=330pt,clip]{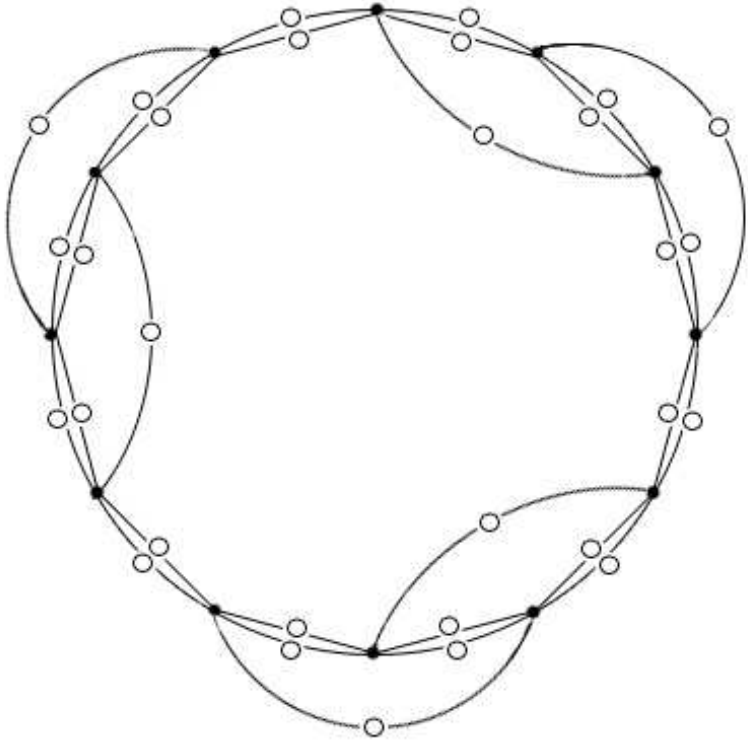}
\caption{The degree 12 invariant $I_{12}$ of a 5-form in 2 variables, represented as a diagram of tensor contraction. Black 5-valent vertices represent tensor $S$, white 2-valent vertices represent tensor $\epsilon$.}
\end{center}
\end{figure}
\paragraph{Invariant form of the Ward identities.}It is possible to deal with non-invariant equations (as we did in the previous section) but it is much more convenient to rewrite the above differential equations in $SL(2)$-invariant form, by summing them with appropriate $S$-dependent coefficients:

\begin{align*}
{\hat O}_0 \ = \ & (2 S_{11111} S_{12222}-8 S_{11112} S_{11222}+6 S_{11122}^2) \left( 2 \dfrac{\partial}{\partial s_{50}} \dfrac{\partial}{\partial s_{14}}-8 \dfrac{\partial}{\partial s_{41}} \dfrac{\partial}{\partial s_{23}}+6 \dfrac{\partial}{\partial s_{32}}\dfrac{\partial}{\partial s_{32}} \right) + \emph{} \\ & 2(S_{11111} S_{22222}-3 S_{11112} S_{12222}+2 S_{11122} S_{11222}) \left( \dfrac{\partial}{\partial s_{05}} \dfrac{\partial}{\partial s_{50}} -3 \dfrac{\partial}{\partial s_{41}} \dfrac{\partial}{\partial s_{14}}+ 2 \dfrac{\partial}{\partial s_{32}} \dfrac{\partial}{\partial s_{23}}\right) + \emph{} \\ & (2 S_{11112} S_{22222}-8 S_{11122} S_{12222}+6 S_{11222}^2) \left(2 \dfrac{\partial}{\partial s_{41}} \dfrac{\partial}{\partial s_{05}}-8 \dfrac{\partial}{\partial s_{32}} \dfrac{\partial}{\partial s_{14}}+6 \dfrac{\partial}{\partial s_{23}}\dfrac{\partial}{\partial s_{23}}\right)
\end{align*}
and
\begin{align*}
{\hat O}_4 \ = \ & P_{11}\big(S\big) \ \left( 2 \dfrac{\partial}{\partial s_{50}} \dfrac{\partial}{\partial s_{14}}-8 \dfrac{\partial}{\partial s_{41}} \dfrac{\partial}{\partial s_{23}}+6 \dfrac{\partial}{\partial s_{32}}\dfrac{\partial}{\partial s_{32}} \right) + \emph{} \\ & 2P_{12}\big(S\big) \ \left(\dfrac{\partial}{\partial s_{05}} \dfrac{\partial}{\partial s_{50}} -3 \dfrac{\partial}{\partial s_{41}} \dfrac{\partial}{\partial s_{14}}+ 2 \dfrac{\partial}{\partial s_{32}} \dfrac{\partial}{\partial s_{23}}\right) + \emph{} \\ & P_{22}\big(S\big) \ \left(2 \dfrac{\partial}{\partial s_{41}} \dfrac{\partial}{\partial s_{05}}-8 \dfrac{\partial}{\partial s_{32}} \dfrac{\partial}{\partial s_{14}}+6 \dfrac{\partial}{\partial s_{23}}\dfrac{\partial}{\partial s_{23}}\right)
\end{align*}
\smallskip\\
where $P$ is a quadratic form with coefficients

\begin{align*}
P_{ab}\big(S\big) \ = \ & S_{a i_2 i_3 i_4 i_5} S_{j_1 j_2 j_3 j_4 j_5} S_{k_1 k_2 k_3 k_4 k_5} S_{l_1 l_2 l_3 l_4 l_5} S_{m_1 m_2 m_3 m_4 m_5} S_{s_1 s_2 s_3 s_4 b} \\ & \\ & \epsilon^{i_2 j_2} \epsilon^{i_3 j_3} \epsilon^{i_4 k_4} \epsilon^{i_5 k_5} \epsilon^{k_1 l_1} \epsilon^{k_2 l_2} \epsilon^{j_4 m_4} \epsilon^{j_5 m_5} \epsilon^{m_1 s_1} \epsilon^{m_2 s_2} \epsilon^{l_3 s_3} \epsilon^{l_4 s_4} \epsilon^{j_1 k_3} \epsilon^{l_5 m_3} \
\end{align*}
\smallskip\\
Operators ${\hat O}_0$ and ${\hat O}_4$ are $SL(2)$-invariant, simply because they are given by diagrams at Fig.11 and Fig.12. The subscripts "0" and "4" stand for the degrees of these operators. In analogy with invariants $I_k$, we denote the invariant differential operators of degree $k$ as ${\hat O}_{k}$.
\begin{figure}[h]
\begin{center}
\includegraphics[totalheight=250pt,clip]{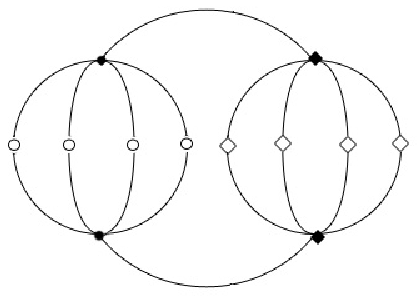}
\caption{Invariant differential operator ${\hat O}_0$ for the case $2|5$, represented as a diagram of tensor contraction. Black 5-valent circles represent tensor $S$, white 2-valent circles represent tensor $\epsilon$, black 5-valent rhombuses represent tensor $\partial / \partial S$, white 2-valent rhombuses represent tensor $\epsilon_*$. }
\bigskip
\bigskip
\includegraphics[totalheight=250pt,clip]{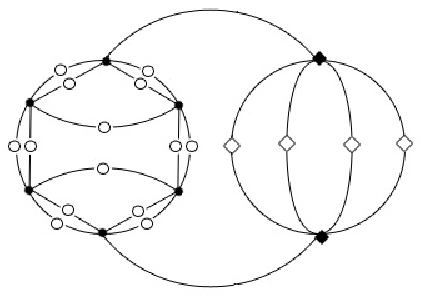}
\caption{Invariant differential operator ${\hat O}_4$ for the case $2|5$, represented as a diagram of tensor contraction. Black 5-valent circles represent tensor $S$, white 2-valent circles represent tensor $\epsilon$, black 5-valent rhombuses represent tensor $\partial / \partial S$, white 2-valent rhombuses represent tensor $\epsilon_*$. }
\end{center}
\end{figure}
\clearpage
Of course, operators ${\hat O}_0$ and ${\hat O}_4$ are not unique -- there are many other invariant differential operators, which annihilate $J_{2|5}$. All such operators contain an $\epsilon$-antisymmetrized combination of two derivatives $\partial / \partial S$, which is exactly the structure of Ward identities. We do not study here the interesting problems of classification of these operators and finding relations between them, because ${\hat O}_0$ and ${\hat O}_4$ are quite enough to find $J_{2|5}$. Let us derive a useful formula for the action of second order operators

$$ \hat O = \sum\limits_{\alpha, \beta} C_{\alpha,\beta} \dfrac{\partial}{\partial S_{\alpha}} \dfrac{\partial}{\partial S_{\beta}} $$
\smallskip\\
By application of the chain rule, we obtain

$$ {\hat O} F\{ I_k \} = \sum\limits_{k} \dfrac{\partial F}{\partial I_k} \sum\limits_{\alpha, \beta} C_{\alpha,\beta} \dfrac{\partial^2 I_k}{\partial S_{\alpha} \partial S_{\beta}} + \sum\limits_{k,m} \dfrac{\partial^2 F}{\partial I_k \partial I_m} \sum\limits_{\alpha, \beta} C_{\alpha,\beta} \dfrac{\partial I_k}{\partial S_{\alpha}} \dfrac{\partial I_m}{\partial S_{\beta}} $$
\smallskip\\
Since

$$\sum\limits_{\alpha, \beta} C_{\alpha,\beta} \dfrac{\partial^2 I_k}{\partial S_{\alpha} \partial S_{\beta}} = {\hat O} I_k$$
\smallskip\\
and

$$\sum\limits_{\alpha, \beta} 2 C_{\alpha,\beta} \dfrac{\partial I_k}{\partial S_{\alpha}} \dfrac{\partial I_m}{\partial S_{\beta}} = {\hat O} \big( I_k I_m \big) - I_k {\hat O} I_m - I_m {\hat O} I_k $$
\smallskip\\
we finally obtain an important formula:

\begin{equation}
\addtolength{\fboxsep}{5pt}
\boxed{
\begin{gathered}
\ \ \ {\hat O} F\{ I_k \} = \sum\limits_{k} \dfrac{\partial F}{\partial I_k} {\hat O} I_k + \dfrac{1}{2} \sum\limits_{k,m} \dfrac{\partial^2 F}{\partial I_k \partial I_m} \Big[ {\hat O} \big( I_k I_m \big) - I_k {\hat O} I_m - I_m {\hat O} I_k \Big] \ \ \
\end{gathered}
}\label{InvariantOperators}
\end{equation}
\smallskip\\
As one can see, to describe the action of ${\hat O}$-operators on arbitrary invariant functions it suffices to calculate the action on all invariants $I_k$ and all products $I_k I_m$. In the present case we have

\[ {\hat O}_{0}
\ \left( \begin{array}{ccc}
\\
I_4 \\
\\
I_8 \\
\\
I_{12} \\
\\
\end{array} \right)
=
\left( \begin{array}{ccc}
\\
\dfrac{264}{25} I_4 \\
\\
\dfrac{2}{25} I_4^2 + \dfrac{294}{25} I_8 \\
\\
\dfrac{12}{25} I_4 I_8 + \dfrac{162}{5} I_{12} \\
\\
\end{array} \right)
\]

{\fontsize{7pt}{0pt}\[ {\hat O}_{0}
\ \left( \begin{array}{ccc}
\\
I_4^2 & I_4 I_8 & I_4 I_{12} \\
\\
I_4 I_8 & I_8^2 & I_8 I_{12} \\
\\
I_4 I_{12} & I_8 I_{12} & I_{12}^2 \\
\\
\end{array} \right)
=
\left( \begin{array}{ccc}
\\
\dfrac{928}{25} I_{4}^2-\dfrac{384}{25} I_{8} & \dfrac{2}{25} I_{4}^3+\dfrac{1166}{25} I_{4} I_{8}+\dfrac{192}{5} I_{12} & \dfrac{12}{25} I_{4}^2 I_{8}+\dfrac{144}{25} I_{8}^2+\dfrac{1794}{25} I_{4} I_{12} \\
\\
\dfrac{2}{25} I_{4}^3+\dfrac{1166}{25} I_{4} I_{8}+\dfrac{192}{5} I_{12} & \dfrac{8}{25} I_{4}^2 I_{8}+\dfrac{1188}{25} I_{8}^2+\dfrac{12}{5} I_{4} I_{12} & \dfrac{9}{25} I_{4} I_{8}^2+\dfrac{12}{25} I_{4}^2 I_{12}+\dfrac{1944}{25} I_{8} I_{12} \\
\\
\dfrac{12}{25} I_{4}^2 I_{8}+\dfrac{144}{25} I_{8}^2+\dfrac{1794}{25} I_{4} I_{12} & \dfrac{9}{25} I_{4} I_{8}^2+\dfrac{12}{25} I_{4}^2 I_{12}+\dfrac{1944}{25} I_{8} I_{12} & -\dfrac{54}{25} I_{8}^3+\dfrac{84}{25} I_{4} I_{8} I_{12}+\dfrac{684}{5} I_{12}^2 \\
\\
\end{array} \right)
\]}
and similarly for the second operator
\[ {\hat O}_{4}
\ \left( \begin{array}{ccc}
\\
I_4 \\
\\
I_8 \\
\\
I_{12} \\
\\
\end{array} \right)
=
\left( \begin{array}{ccc}
\\
-\dfrac{264}{25} I_{8} \\
\\
-\dfrac{2}{25} I_{4} I_{8}+\dfrac{588}{25} I_{12} \\
\\
\dfrac{363}{50} I_{8}^2-\dfrac{153}{25} I_{4} I_{12} \\
\\
\end{array} \right)
\]

{\fontsize{7pt}{0pt}\[ {\hat O}_{4}
\ \left( \begin{array}{ccc}
\\
I_4^2 & I_4 I_8 & I_4 I_{12} \\
\\
I_4 I_8 & I_8^2 & I_8 I_{12} \\
\\
I_4 I_{12} & I_8 I_{12} & I_{12}^2 \\
\\
\end{array} \right)
=
\left( \begin{array}{ccc}
\\
-\dfrac{928}{25} I_{4} I_{8}-\dfrac{768}{25} I_{12} & -\dfrac{2}{25} I_{4}^2 I_{8}-\dfrac{584}{25} I_{8}^2+\dfrac{524}{25} I_{4} I_{12} & \dfrac{363}{50} I_{4} I_{8}^2-\dfrac{153}{25} I_{4}^2 I_{12}-\dfrac{696}{25} I_{8} I_{12} \\
\\
-\dfrac{2}{25} I_{4}^2 I_{8}-\dfrac{584}{25} I_{8}^2+\dfrac{524}{25} I_{4} I_{12} & \dfrac{1}{25} I_{4} I_{8}^2-\dfrac{22}{25} I_{4}^2 I_{12}+\dfrac{2376}{25} I_{8} I_{12} & \dfrac{603}{50} I_{8}^3-\dfrac{291}{25} I_{4} I_{8} I_{12}+\dfrac{1188}{25} I_{12}^2 \\
\\
\dfrac{363}{50} I_{4} I_{8}^2-\dfrac{153}{25} I_{4}^2 I_{12}-\dfrac{696}{25} I_{8} I_{12} & \dfrac{603}{50} I_{8}^3-\dfrac{291}{25} I_{4} I_{8} I_{12}+\dfrac{1188}{25} I_{12}^2 & \dfrac{129}{5} I_{8}^2 I_{12}-\dfrac{606}{25} I_{4} I_{12}^2 \\
\\
\end{array} \right)
\]}
Applying (\ref{InvariantOperators}), we obtain

$$ {\hat O}_0 F \big( I_4, I_8, I_{12} \big) = \dfrac{264}{25} I_4 \dfrac{\partial F}{\partial I_4} + \left( \dfrac{2}{25} I_4^2 + \dfrac{294}{25} I_8 \right) \dfrac{\partial F}{\partial I_8} + \left( \dfrac{12}{25} I_4 I_8 + \dfrac{162}{5} I_{12} \right) \dfrac{\partial F}{\partial I_{12}} $$

$$ + \left( 8 I_{4}^2-\dfrac{192}{25} I_{8} \right) \dfrac{\partial^2 F}{\partial I_4^2 } + \left(\dfrac{2}{25} I_{4}^2 I_{8}+12 I_{8}^2+\dfrac{6}{5} I_{4} I_{12}\right) \dfrac{\partial^2 F}{\partial I_8^2 } + \left( -\dfrac{27}{25} I_{8}^3+\dfrac{6}{5} I_{4} I_{8} I_{12}+36 I_{12}^2 \right) \dfrac{\partial^2 F}{\partial I_{12}^2 } $$

$$ + \left( \dfrac{608}{25} I_{4} I_{8}+\dfrac{192}{5} I_{12} \right) \dfrac{\partial^2 F}{\partial I_4 \partial I_8 } + \left( \dfrac{144}{25} I_{8}^2+\dfrac{144}{5} I_{4} I_{12} \right) \dfrac{\partial^2 F}{\partial I_4 \partial I_{12} } + \left( -\dfrac{3}{25} I_{4} I_{8}^2+\dfrac{2}{5} I_{4}^2 I_{12}+\dfrac{168}{5} I_{8} I_{12} \right) \dfrac{\partial^2 F}{\partial I_8 \partial I_{12} } $$
\smallskip\\
and similarly for the second operator

$$ {\hat O}_4 F \big( I_4, I_8, I_{12} \big) = -\dfrac{264}{25} I_{8} \dfrac{\partial F}{\partial I_4} + \left( -\dfrac{2}{25} I_{4} I_{8}+\dfrac{588}{25} I_{12} \right) \dfrac{\partial F}{\partial I_8} + \left( \dfrac{363}{50} I_{8}^2-\dfrac{153}{25} I_{4} I_{12} \right) \dfrac{\partial F}{\partial I_{12}} $$

$$ + \left( -8 I_{4} I_{8}-\dfrac{384}{25} I_{12} \right) \dfrac{\partial^2 F}{\partial I_4^2 } + \left(\dfrac{1}{10} I_{4} I_{8}^2-\dfrac{11}{25} I_{4}^2 I_{12}+24 I_{8} I_{12}\right) \dfrac{\partial^2 F}{\partial I_8^2 } + \left( \dfrac{141}{25} I_{8}^2 I_{12}- 6 I_{4} I_{12}^2 \right) \dfrac{\partial^2 F}{\partial I_{12}^2 } $$

$$ + \left( -\dfrac{64}{5} I_{8}^2-\dfrac{64}{25} I_{4} I_{12}\right) \dfrac{\partial^2 F}{\partial I_4 \partial I_8 } - \dfrac{432}{25} I_{8} I_{12} \dfrac{\partial^2 F}{\partial I_4 \partial I_{12} } + \left( \dfrac{24}{5} I_{8}^3-\dfrac{136}{25} I_{4} I_{8} I_{12}+24 I_{12}^2 \right) \dfrac{\partial^2 F}{\partial I_8 \partial I_{12} } $$
\smallskip\\
For the integral discriminant, both expressions vanish. Substituting the ansatz (\ref{Ansatz25}) and making the necessary algebraic transformations, we obtain two differential equations on the function $G(u,v)$: the first

\begin{align}
& \nonumber 50(-1+64u)(u+6u^2+15 v) \dfrac{\partial^2 G}{\partial u^2}+(75 u^2+72000 v^2+57600 v u^2+600 v u+7200 u^3-250 v) \dfrac{\partial^2 G}{\partial u \partial v} + \emph{} \\ & \nonumber (675 u^3-13500 v^2+10800 v u^2-750 v u+43200 u v^2) \dfrac{\partial^2 G}{\partial v^2} + (50400 v+30720 u^2-50+5770 u) \dfrac{\partial G}{\partial u} + \emph{} \\ & \nonumber (-300 u+60480 v u+11160 u^2-7650 v) \dfrac{\partial G}{\partial v}+(528 u+110) G = 0
\end{align}
\smallskip\\
and the second
\begin{align}
& \nonumber 25(-1+64u)(5u^2+48uv-22v) \dfrac{\partial^2 G}{\partial u^2}+(230400 u v^2-39600 v^2-6000 u^3+28800 v u^2+6800 v u) \dfrac{\partial^2 G}{\partial u \partial v}+ \emph{} \\ & \nonumber (172800 v^3+7500 v^2+25200 u v^2-7050 v u^2) \dfrac{\partial^2 G}{\partial v^2}+(-36120 v+4000 u^2+122880 v u+100 u) \dfrac{\partial G}{\partial u}+ \emph{} \\ & \nonumber (241920 v^2+19440 v u-9075 u^2+7650 v) \dfrac{\partial G}{\partial v}+(-220 u+2112 v) G(u,v) = 0
\end{align}
where $u = I_8/I_4^2$ and $v = I_{12}/I_4^3$. These two linear differential equations in two variables constitute the invariant form of $n = 2, r = 5$ Ward identities. Integral discriminant is found as the solution of this system.

\paragraph{Solution.}Having linear differential equations, it is easy to solve them in series: if one puts

$$ G(u,v) = \sum\limits_{i,j} c_{ij} u^i v^j $$
\smallskip\\
then one finds, after some algebraic transformations,

$$ c_{ij} = {\rm const} \cdot \dfrac{1}{i!j!} \cdot 16^i \left(\dfrac{128}{3}\right)^j \cdot \dfrac{\Gamma\left(\dfrac{3}{10} + i + j \right) \Gamma\left(\dfrac{1}{10} + 2i + 3j \right) \Gamma\left(\dfrac{1}{10} + j \right)}{\Gamma\left(\dfrac{2}{5} + i + 2j \right)\Gamma\left(\dfrac{3}{5} + i + 2j \right) } $$
\smallskip\\
i.e. there is unique series solution. In this way one obtains the first branch of the integral discriminant $J_{2|5}$:

\begin{equation}
\addtolength{\fboxsep}{5pt}
\!\boxed{
\begin{gathered}
J^{(1)}_{2|5}\big(S\big) = I_4^{-1/10} \sum\limits_{i,j} \dfrac{\Gamma\left(\dfrac{3}{10} + i + j \right) \Gamma\left(\dfrac{1}{10} + 2i + 3j \right) \Gamma\left(\dfrac{1}{10} + j \right)}{\Gamma\left(\dfrac{2}{5} + i + 2j \right)\Gamma\left(\dfrac{3}{5} + i + 2j \right) } \cdot \dfrac{1}{i!j!} \cdot \left(\dfrac{16I_8}{I_4^2}\right)^i \left(\dfrac{128I_{12}}{3I_{4}^3}\right)^j
\end{gathered}
}\label{I25}
\end{equation}
\smallskip\\
This answer is interesting: the function appears to be of generalised hypergeometric type \cite{genhypergeom}, since its coefficients $c_{ij}$ are ratios of $\Gamma$-functions, depending on linear combinations of indices $i$ and $j$. However, these series have a rather small convergence radius. Already the subsequence with $i=j$ behaves as

$$\sum\limits_{j} \dfrac{\Gamma\left(\dfrac{3}{10} + 2j \right) \Gamma\left(\dfrac{1}{10} + 5j \right) \Gamma\left(\dfrac{1}{10} + j \right)}{\Gamma\left(\dfrac{2}{5} + 3j \right)\Gamma\left(\dfrac{3}{5} + 3j \right) } \left( \dfrac{2^{11}}{3} uv \right)^j \sim \sum\limits_{j} \dfrac{1}{j^2} \left( \dfrac{2^{13} 5^5}{3^7} uv \right)^j $$
\smallskip\\
and diverges when the combination in the last brackets exceeds unity. For analytical continuation, one better substitutes this series by its integral representation.

\paragraph{Hypergeometric integral.} Applying the formula (\ref{BFunction}) twice, we get

{\fontsize{9pt}{0pt}$$\dfrac{\Gamma\left(\dfrac{3}{10} + i + j \right) \Gamma\left(\dfrac{1}{10} + 2i + 3j \right) \Gamma\left(\dfrac{1}{10} + j \right)}{\Gamma\left(\dfrac{2}{5} + i + 2j \right)\Gamma\left(\dfrac{3}{5} + i + 2j \right) } = \dfrac{1}{\Gamma\left( \dfrac{1}{2} - i - j \right)} \int\limits_{0}^{1} \int\limits_{0}^{1} dt ds \ \dfrac{ t^{-7/10} (s-ts)^{-9/10} }{\sqrt{1 - s}} \ \left(\dfrac{ts^2}{1-s}\right)^i \left( \dfrac{t(1-t)s^3}{1-s} \right)^j  $$}
\smallskip\\
The sum over $i$ and $j$ is calculated, using

$$ \sum\limits_{i,j} \dfrac{1}{\Gamma\left( \dfrac{1}{2} - i - j \right)} \dfrac{A^i}{i!} \dfrac{B^j}{j!} = \dfrac{1}{\sqrt{1 + A + B}} $$
\smallskip\\
and we obtain the following integral representation:

\begin{align}
G(u,v) = \int\limits_{0}^{1}\int\limits_{0}^{1} dt ds \cdot \dfrac{t^{-7/10} s^{-9/10} (1-t)^{-9/10}}{\sqrt{3-3 s+48 u t s^2+128 v s^3 t-128 v s^3 t^2}}
\label{Integral25}
\end{align}
\smallskip\\
Accordingly, for $J_{2|5}$ we have

\begin{equation}
\addtolength{\fboxsep}{5pt}
\boxed{
\begin{gathered}
J^{(1)}_{2|5}\big(S\big) = I_{4}^{7/5} \int\limits_{0}^{1}\int\limits_{0}^{1} dt ds \cdot \dfrac{t^{-7/10} s^{-9/10} (1-t)^{-9/10}}{\sqrt{3 I_{4}^3 - 3 I_{4}^3 s + 48 I_4 I_8 t s^2 + 128 I_{12} s^3 t-128 I_{12} s^3 t^2}}
\end{gathered}
}\label{integ25}
\end{equation}
\smallskip\\
Other branches of $J_{2|5}$ can be obtained by various transformations of variables $(u,v)$ and integration variables $(t,s)$, which leave the boundary region intact ("modular" transformations). We do not list them in this paper.

\paragraph{Singularities.}One of the benefits of integral representation is the possibility to analyse the singularities. Singularities of the integral (\ref{Integral25}) come from zeroes of the polynomial

$$ P(s,t)=3-3 s+48 u t s^2+128 v s^3 t-128 v s^3 t^2$$
\smallskip\\
which stands in the denominator. Ordinary zeroes (say, in $t$) are inessential, already because the singularity $dt/\sqrt{t-t_0}$ is integrable, and -- more important -- because the integration contour can be deformed and taken away from the singularity. This can not be done if the two roots coincide, i.e. if discriminant of $P(t)$ is vanishing. Then what matters is integration over $s$. If

$$\mbox{ Disc}_t(P(s,t)) \sim (t_1(s)-t_2(s))^2$$
\smallskip\\
has a simple zero at $s=s_0$, then

$$P(s,t) \sim (t-t_0)^2 + (t_1-t_2)^2 \sim (t-t_0)^2 + (s-s_0)$$
\smallskip\\
and the integration contour for $s$ can be taken away from the singularity. More serious is the case when discriminant has double zero, i.e. when repeated discriminant vanishes,

$$\mbox{ Disc}_s\Big(\mbox{ Disc}_t(P(s,t))\Big) = 0$$
\smallskip\\
Then contour can not be deformed, neither in $t$ nor in $s$. However, in the
vicinity of such point

$$P(s,t) \sim (t-t_0)^2+(s-s_0)^2$$
\smallskip\\
and this singularity, though unavoidable, is integrable because of the square root and because the integral is two-dimensional. This is most simply expressed in polar coordinates:

$$ \dfrac{ds dt}{\sqrt{(t-t_0)^2 + (s-s_0)^2}} \sim dr d\phi $$
\smallskip\\
Moreover, even if repeated discriminant has zeroes of higher order, $P(s,t) \sim  (t-t_0)^2 + (s-s_0)^k$, the singularity in the integral does not arise.
\begin{figure}[t]
\begin{center}
\includegraphics[totalheight=200pt,clip]{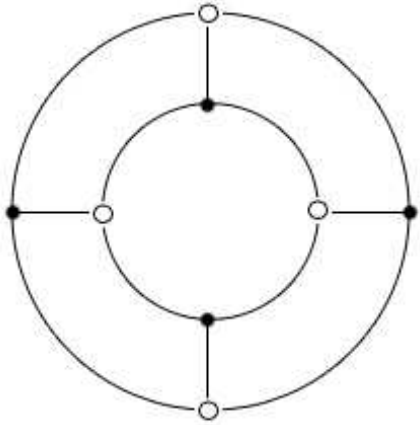}
\caption{The degree 4 invariant $I_{4}$ of a 3-form in 3 variables, represented as a diagram of tensor contraction. Black 3-valent vertices represent tensor $S$, white 3-valent vertices represent tensor $\epsilon$. }
\end{center}
\end{figure}

Thus the only remaining source of singularity is when it
occurs at the ends of integration segments (at the boundary
of integration domain -- because the contour can not
be deformed to avoid these boundaries), i.e. when
$s$ or $t$ equals $0$ or $1$. Given the shape of $P(s,t)$,
of these the only significant one is at $t=1$, when $P(s,1) = 3-3s+48us^2$. Remaining integral over $s$ becomes singular when this expression has a double zero:

$$ \mbox{ Disc}_s\Big( 3-3s+48us^2 \Big) = 1-64u=0$$
\smallskip\\
i.e. when $u=1/64$. This point corresponds to

$$ \dfrac{64I_{8}}{I_{4}^2} = 1$$
\smallskip\\
which is just the discriminant's vanishing locus $ I_{4}^2 - 64I_{8} = D_{2|5} = 0$. We encounter once again the same relation between integral and algebraic discriminants: the latter controls essential singularities of the former. Note, that the double zero occurs at the point $s = 2$, which lies beyond the integration domain. For this reason, the singularity is rather soft.

\subsection{The case of $J_{3|3}$}

Our final example is a $3$-form in $3$ variables, which can be written as

$$ S(x,y) = S_{111} x^3 + 3 S_{112} x^2 y + 3 S_{113} x^2 z + S_{222} y^3 + 3 S_{122} xy^2 + 3 S_{223} y^2z + S_{333} z^3 + 3 S_{133} xz^2 + 3 S_{233} yz^2 + 6 S_{123} xyz  $$
\paragraph{Invariants.}By dimension counting, there are two elementary invariants in this case. They have degrees $4$ and $6$, denoted as $I_{4}, I_{6}$ and given by diagrams at Fig. 13, Fig. 14. Looking at the diagrams, it is straightforward to write the algebraic expressions for $I_{4}$ and $I_{6}$:

$$
I_4 = S_{i_1 i_2 i_3} S_{j_1 j_2 j_3} S_{k_1 k_2 k_3} S_{l_1 l_2 l_3} \epsilon^{i_1 j_1 k_1} \epsilon^{i_2 j_2 l_2} \epsilon^{i_3 k_3 l_3} \epsilon^{l_1 k_2 j_3}
$$

$$
I_6 = S_{i_1 i_2 i_3} S_{j_1 j_2 j_3} S_{k_1 k_2 k_3} S_{l_1 l_2 l_3} S_{m_1 m_2 m_3} S_{s_1 s_2 s_3} \epsilon^{i_1 k_1 l_1}\epsilon^{i_2 j_2 s_2}\epsilon^{j_1 k_2 m_1}\epsilon^{l_2 m_2 k_3}\epsilon^{m_3 s_3 j_3}\epsilon^{l_3 i_3 s_1}
$$
\smallskip\\
Evaluating these sums, one gets the following explicit formulas for $I_4, I_6$:

\begin{figure}[t]
\begin{center}
\includegraphics[totalheight=230pt,clip]{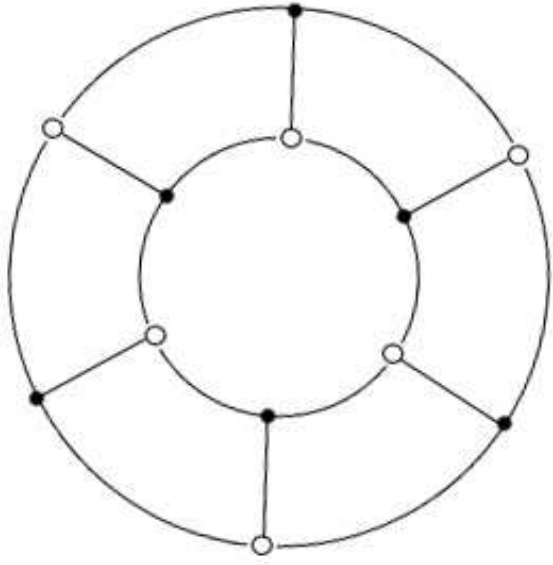}
\caption{The degree 6 invariant $I_{6}$ of a 3-form in 3 variables, represented as a diagram of tensor contraction. Black 3-valent vertices represent tensor $S$, white 3-valent vertices represent tensor $\epsilon$. }
\end{center}
\end{figure}

\begin{center}
$ I_{4} = 6 S_{123}^4-12 S_{122} S_{123}^2 S_{133}+6 S_{122}^2 S_{133}^2+6 S_{113} S_{123} S_{133} S_{222}-12 S_{113} S_{123}^2 S_{223}-6 S_{113} S_{122} S_{133} S_{223}+18 S_{113} S_{122} S_{123} S_{233}-6 S_{113} S_{122}^2 S_{333}+6 S_{113}^2 S_{223}^2-6 S_{113}^2 S_{222} S_{233}-6 S_{112} S_{133}^2 S_{222}+18 S_{112} S_{123} S_{133} S_{223}-12 S_{112} S_{123}^2 S_{233}-6 S_{112} S_{122} S_{133} S_{233}+6 S_{112} S_{122} S_{123} S_{333}-6 S_{112} S_{113} S_{223} S_{233}+6 S_{112} S_{113} S_{222} S_{333}+6 S_{112}^2 S_{233}^2-6 S_{112}^2 S_{223} S_{333}-6 S_{111} S_{133} S_{223}^2+6 S_{111} S_{133} S_{222} S_{233}+6 S_{111} S_{123} S_{223} S_{233}-6 S_{111} S_{123} S_{222} S_{333}-6 S_{111} S_{122} S_{233}^2+6 S_{111} S_{122} S_{223} S_{333}$
\end{center}

\begin{center}
$ I_6 = 48 S_{123}^6-144 S_{122} S_{123}^4 S_{133}+144 S_{122}^2 S_{123}^2 S_{133}^2-48 S_{122}^3 S_{133}^3+72 S_{113} S_{123}^3 S_{133} S_{222}-144 S_{113} S_{123}^4 S_{223}-72 S_{113} S_{122} S_{123} S_{133}^2 S_{222}+72 S_{113} S_{122} S_{123}^2 S_{133} S_{223}+216 S_{113} S_{122} S_{123}^3 S_{233}+72 S_{113} S_{122}^2 S_{133}^2 S_{223}-216 S_{113} S_{122}^2 S_{123} S_{133} S_{233}-72 S_{113} S_{122}^2 S_{123}^2 S_{333}+72 S_{113} S_{122}^3 S_{133} S_{333}+18 S_{113}^2 S_{133}^2 S_{222}^2-72 S_{113}^2 S_{123} S_{133} S_{222} S_{223}+144 S_{113}^2 S_{123}^2 S_{223}^2-72 S_{113}^2 S_{123}^2 S_{222} S_{233}+72 S_{113}^2 S_{122} S_{133} S_{223}^2-36 S_{113}^2 S_{122} S_{133} S_{222} S_{233}-216 S_{113}^2 S_{122} S_{123} S_{223} S_{233}+144 S_{113}^2 S_{122} S_{123} S_{222} S_{333}+162 S_{113}^2 S_{122}^2 S_{233}^2-144 S_{113}^2 S_{122}^2 S_{223} S_{333}-48 S_{113}^3 S_{223}^3+72 S_{113}^3 S_{222} S_{223} S_{233}-24 S_{113}^3 S_{222}^2 S_{333}-72 S_{112} S_{123}^2 S_{133}^2 S_{222}+216 S_{112} S_{123}^3 S_{133} S_{223}-144 S_{112} S_{123}^4 S_{233}+72 S_{112} S_{122} S_{133}^3 S_{222}-216 S_{112} S_{122} S_{123} S_{133}^2 S_{223}+72 S_{112} S_{122} S_{123}^2 S_{133} S_{233}+72 S_{112} S_{122} S_{123}^3 S_{333}+72 S_{112} S_{122}^2 S_{133}^2 S_{233}-72 S_{112} S_{122}^2 S_{123} S_{133} S_{333}-36 S_{112} S_{113} S_{133}^2 S_{222} S_{223}-216 S_{112} S_{113} S_{123} S_{133} S_{223}^2+360 S_{112} S_{113} S_{123} S_{133} S_{222} S_{233}+72 S_{112} S_{113} S_{123}^2 S_{223} S_{233}-216 S_{112} S_{113} S_{123}^2 S_{222} S_{333}+36 S_{112} S_{113} S_{122} S_{133} S_{223} S_{233}-108 S_{112} S_{113} S_{122} S_{133} S_{222} S_{333}-216 S_{112} S_{113} S_{122} S_{123} S_{233}^2+360 S_{112} S_{113} S_{122} S_{123} S_{223} S_{333}-36 S_{112} S_{113} S_{122}^2 S_{233} S_{333}+72 S_{112} S_{113}^2 S_{223}^2 S_{233}-144 S_{112} S_{113}^2 S_{222} S_{233}^2+72 S_{112} S_{113}^2 S_{222} S_{223} S_{333}+162 S_{112}^2 S_{133}^2 S_{223}^2-144 S_{112}^2 S_{133}^2 S_{222} S_{233}-216 S_{112}^2 S_{123} S_{133} S_{223} S_{233}+144 S_{112}^2 S_{123} S_{133} S_{222} S_{333}+144 S_{112}^2 S_{123}^2 S_{233}^2-72 S_{112}^2 S_{123}^2 S_{223} S_{333}+72 S_{112}^2 S_{122} S_{133} S_{233}^2-36 S_{112}^2 S_{122} S_{133} S_{223} S_{333}-72 S_{112}^2 S_{122} S_{123} S_{233} S_{333}+18 S_{112}^2 S_{122}^2 S_{333}^2+72 S_{112}^2 S_{113} S_{223} S_{233}^2-144 S_{112}^2 S_{113} S_{223}^2 S_{333}+72 S_{112}^2 S_{113} S_{222} S_{233} S_{333}-48 S_{112}^3 S_{233}^3+72 S_{112}^3 S_{223} S_{233} S_{333}-24 S_{112}^3 S_{222} S_{333}^2-24 S_{111} S_{133}^3 S_{222}^2+144 S_{111} S_{123} S_{133}^2 S_{222} S_{223}-72 S_{111} S_{123}^2 S_{133} S_{223}^2-216 S_{111} S_{123}^2 S_{133} S_{222} S_{233}+72 S_{111} S_{123}^3 S_{223} S_{233}+120 S_{111} S_{123}^3 S_{222} S_{333}-144 S_{111} S_{122} S_{133}^2 S_{223}^2+72 S_{111} S_{122} S_{133}^2 S_{222} S_{233}+360 S_{111} S_{122} S_{123} S_{133} S_{223} S_{233}-72 S_{111} S_{122} S_{123} S_{133} S_{222} S_{333}-72 S_{111} S_{122} S_{123}^2 S_{233}^2-216 S_{111} S_{122} S_{123}^2 S_{223} S_{333}-144 S_{111} S_{122}^2 S_{133} S_{233}^2+72 S_{111} S_{122}^2 S_{133} S_{223} S_{333}+144 S_{111} S_{122}^2 S_{123} S_{233} S_{333}-24 S_{111} S_{122}^3 S_{333}^2+72 S_{111} S_{113} S_{133} S_{223}^3-108 S_{111} S_{113} S_{133} S_{222} S_{223} S_{233}+36 S_{111} S_{113} S_{133} S_{222}^2 S_{333}-72 S_{111} S_{113} S_{123} S_{223}^2 S_{233}+144 S_{111} S_{113} S_{123} S_{222} S_{233}^2-72 S_{111} S_{113} S_{123} S_{222} S_{223} S_{333}-36 S_{111} S_{113} S_{122} S_{223} S_{233}^2+72 S_{111} S_{113} S_{122} S_{223}^2 S_{333}-36 S_{111} S_{113} S_{122} S_{222} S_{233} S_{333}-36 S_{111} S_{112} S_{133} S_{223}^2 S_{233}+72 S_{111} S_{112} S_{133} S_{222} S_{233}^2-36 S_{111} S_{112} S_{133} S_{222} S_{223} S_{333}-72 S_{111} S_{112} S_{123} S_{223} S_{233}^2+144 S_{111} S_{112} S_{123} S_{223}^2 S_{333}-72 S_{111} S_{112} S_{123} S_{222} S_{233} S_{333}+72 S_{111} S_{112} S_{122} S_{233}^3-108 S_{111} S_{112} S_{122} S_{223} S_{233} S_{333}+36 S_{111} S_{112} S_{122} S_{222} S_{333}^2+18 S_{111}^2 S_{223}^2 S_{233}^2-24 S_{111}^2 S_{223}^3 S_{333}-24 S_{111}^2 S_{222} S_{233}^3+36 S_{111}^2 S_{222} S_{223} S_{233} S_{333}-6 S_{111}^2 S_{222}^2 S_{333}^2$ \smallskip\\\end{center}
The algebraic discriminant $D_{3|3}$, just like any other $SL(3)$-invariant function of $S$, is a function of $I_4$ and $I_6$:

\begin{align}
D_{3|3} = 32 I_4^3 + 3 I_6^2
\label{D33}
\end{align}
\smallskip\\
When expanded, discriminant $D_{3|3}$ contains 2040 monomials. See the appendix of the book version of \cite{nolal}, where it is written explicitly for curiosity. Formula (\ref{D33}) is a remarkably concise expression of this disriminant through a pair of invariants, given by beautiful diagrams Fig.13 and Fig.14. It is interesting to extend this type of formulas -- eqs. (\ref{D23}), (\ref{D24}), (\ref{D25}) and (\ref{D33}) -- to higher $n$ and $r$.

\paragraph{Integral discriminant.}Similarly, the integral discriminant is a function of $I_{4}, I_{6}$:

$$J_{3|3} = F\big( I_4, I_6 \big)$$
\smallskip\\
where the function $F$ is to be determined. The homogeneity condition does not allow to find $F$ unambigously, it constrains only the overall scaling factor, but not the non-trivial dependence on the ratio of invariants:

\begin{align}
F\big( I_4, I_6 \big) = I_{4}^{-1/4} \ G\left( \dfrac{I_6^2}{I_4^3} \right)
\label{Ansatz33}
\end{align}
\smallskip\\
To find the function $G(z)$ in this ansatz, we need to use the Ward identities (\ref{Ward}).

\paragraph{Ward identities.} Applied to the present case of $n = 3, r = 3$, the system (\ref{Ward}) takes the form

\[
\begin{array}{ccc}
\left( \dfrac{\partial^2}{\partial s_{300} \partial s_{102}} - \dfrac{\partial^2}{\partial s_{201} \partial s_{201} } \right) J_{3|3} = 0 \\
\\
\left( \dfrac{\partial^2}{\partial s_{300} \partial s_{111}} - \dfrac{\partial^2}{\partial s_{201} \partial s_{210} } \right) J_{3|3} = 0 \\
\\
\left( \dfrac{\partial^2}{\partial s_{300} \partial s_{003}} - \dfrac{\partial^2}{\partial s_{201} \partial s_{102} } \right) J_{3|3} = 0 \\
\\
\left( \dfrac{\partial^2}{\partial s_{300} \partial s_{012}} - \dfrac{\partial^2}{\partial s_{201} \partial s_{111} } \right) J_{3|3} = 0 \\
\\
\left( \dfrac{\partial^2}{\partial s_{300} \partial s_{012}} - \dfrac{\partial^2}{\partial s_{210} \partial s_{102} } \right) J_{3|3} = 0 \\
\\
\end{array}\ \ \
\begin{array}{ccc}
\left( \dfrac{\partial^2}{\partial s_{300} \partial s_{120}} - \dfrac{\partial^2}{\partial s_{210} \partial s_{210} } \right) J_{3|3} = 0 \\
\\
\left( \dfrac{\partial^2}{\partial s_{300} \partial s_{111}} - \dfrac{\partial^2}{\partial s_{210} \partial s_{201} } \right) J_{3|3} = 0 \\
\\
\left( \dfrac{\partial^2}{\partial s_{300} \partial s_{030}} - \dfrac{\partial^2}{\partial s_{210} \partial s_{120} } \right) J_{3|3} = 0 \\
\\
\left( \dfrac{\partial^2}{\partial s_{300} \partial s_{021}} - \dfrac{\partial^2}{\partial s_{210} \partial s_{111} } \right) J_{3|3} = 0 \\
\\
\left( \dfrac{\partial^2}{\partial s_{300} \partial s_{021}} - \dfrac{\partial^2}{\partial s_{201} \partial s_{120} } \right) J_{3|3} = 0 \\
\\
\end{array}
\]
where $s$-parameters and $S$-parameters are related by

$$ s_{300} = S_{111}, \ \ s_{210} = 3 S_{112}, \ \ s_{201} = 3 S_{113}, \ \ s_{030} = S_{222}, \ \ s_{120} = 3 S_{122}, \ \ s_{021} = 3 S_{223}$$

$$ s_{003} = S_{333}, \ \ s_{102} = 3 S_{133}, \ \ s_{012} = 3 S_{233}, \ \ s_{111} = 6 S_{123} $$
\smallskip\\
Particular equations in this system are, of course, not $SL(3)$ invariant.

\paragraph{Invariant form of Ward identities.} To rewrite the above differential equations in $SL(3)$-invariant form, we use again the method of invariant differential operators -- sum the Ward operators with appropriate $S$-dependent coefficients to form an invariant operator ${\hat O}_4$, given by the diagram at Fig.15. The antisymmetrized combination of two derivatives (two black rhombuses) which is a part of this diagram, \begin{figure}[h]
\begin{center}
\includegraphics[totalheight=200pt,clip]{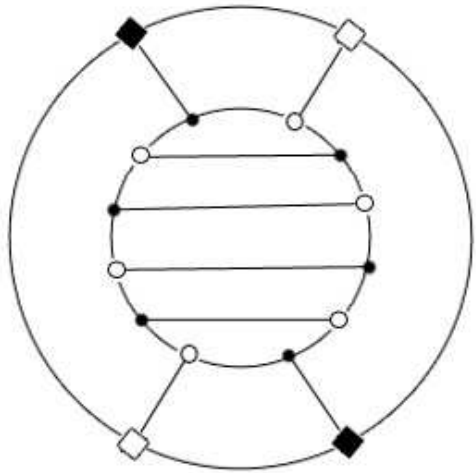}
\caption{Invariant differential operator ${\hat O}_4$ for the case $3|3$, represented as a diagram of tensor contraction. Black 3-valent vertices represent tensor $S$, white 3-valent vertices represent tensor $\epsilon$, black 3-valent rhombuses represent tensor $\partial / \partial S$, white 3-valent rhombuses represent tensor $\epsilon_*$.}
\end{center}
\end{figure} ensures that ${\hat O}_4$ is a linear combination of Ward operators and annihilates $J_{3|3}$. Operator ${\hat O}_4$ contains $4$ horizontal lines in the inner circle and belongs to an infinite family of operators with $2p$ horizontal lines in the inner circle:

$${\hat O}_{2p}, \ \ \ p = 1,2,3,\ldots$$
\smallskip\\
of which the simplest are ${\hat O}_0$ and ${\hat O}_2$, given by diagrams at Fig.16. However, operators ${\hat O}_0$ and ${\hat O}_2$ are too simple: in fact, they do not constrain $J_{3|3}$ at all. Thus ${\hat O}_4$ is the main operator we use in this section.

\begin{figure}[h]
\begin{center}
\includegraphics[totalheight=200pt,clip]{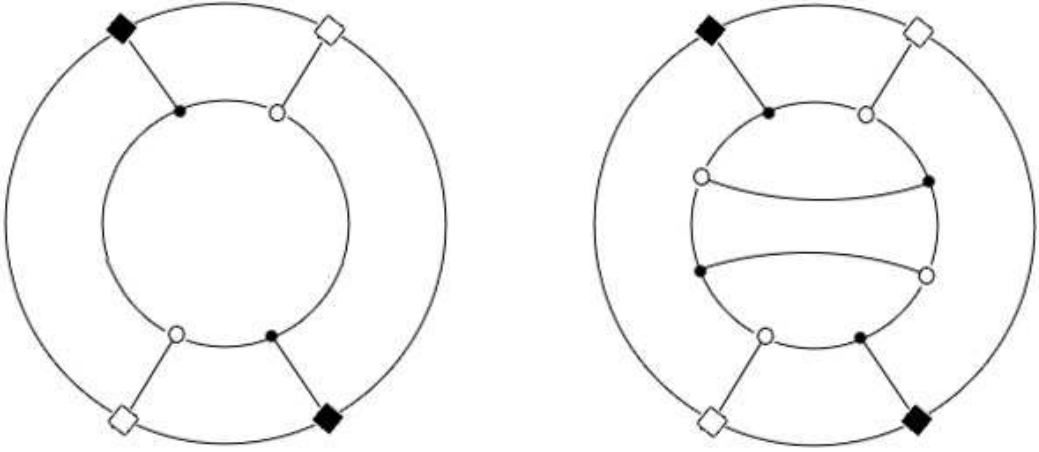}
\caption{Invariant differential operators ${\hat O}_0$ and ${\hat O}_2$ for the case $3|3$, represented as diagrams of tensor contraction. Black 3-valent vertices represent tensor $S$, white 3-valent vertices represent tensor $\epsilon$, black 3-valent rhombuses represent tensor $\partial / \partial S$, white 3-valent rhombuses represent tensor $\epsilon_*$. These operators cannot be used to find $J_{3|3}$, because they do not give any non-trivial equation on it.}
\end{center}
\end{figure}

The action of ${\hat O}_4$ on invariants and their products is given by

\[ {\hat O}_{4}
\ \left( \begin{array}{ccc}
\\
I_4 \\
\\
I_6 \\
\\
\end{array} \right)
=
\left( \begin{array}{ccc}
\\
-\dfrac{140}{9} I_4^2 \\
\\
-\dfrac{98}{3} I_4 I_6 \\
\\
\end{array} \right)
\]

\[ {\hat O}_{4}
\ \left( \begin{array}{ccc}
\\
I_4^2 & I_4 I_6 \\
\\
I_4 I_6 & I_6^2  \\
\\
\end{array} \right)
=
\left( \begin{array}{ccc}
\\
\dfrac{1}{3} I_{6}^2 - \dfrac{472}{9} I_{4}^3 & -\dfrac{770}{9} I_4^2 I_6 \\
\\
-\dfrac{770}{9} I_4^2 I_6 & \dfrac{256}{3} I_{4}^4 - \dfrac{340}{3} I_{6}^2 I_4\\
\\
\end{array} \right)
\]
Applying (\ref{InvariantOperators}), we obtain

$$ {\hat O}_4 F \big( I_4, I_6 \big) = -\dfrac{140}{9} I_4^2 \dfrac{\partial F}{\partial I_4} - \dfrac{98}{3} I_4 I_6 \dfrac{\partial F}{\partial I_6} + \left( \dfrac{1}{6} I_6^2 - \dfrac{32}{9} I_{4}^3 \right) \dfrac{\partial^2 F}{\partial I_4^2} - \dfrac{112}{3} I_4^2 I_6 \dfrac{\partial^2 F}{\partial I_4 \partial I_6} + \left( \dfrac{128}{3} I_4^4 - 24 I_4 I_6^2 \right) \dfrac{\partial^2 F}{\partial I_6^2} $$
\smallskip\\
For the integral discriminant, this expression vanishes. Substituting the ansatz (\ref{Ansatz33}) and making the necessary algebraic transformations, we obtain a differential equation on the function $G(z)$:

\begin{align}
( 144 z^2 + 1536 z ) \dfrac{\partial^2 G(z)}{\partial z^2} + \left( 216 z + 768 \right) \dfrac{\partial G(z)}{\partial z} + 5 G(z) = 0
\label{33hypergeom}
\end{align}
\smallskip\\
which is nothing but Gauss hypergeometric equation

$$
t(1 - t) \dfrac{\partial^2 G(t)}{\partial t^2} + \left( c - (a + b + 1) t \right) \dfrac{\partial G(t)}{\partial t} - ab G(t) = 0
$$
\smallskip\\
with $a = 1/12$, $b = 5/12$, $c = 1/2$ and $t = -3/32z$. This is the invariant form of $n = 3, r = 3$ Ward identities.

\paragraph{Solution.}In terms of the Gauss hypergeometric function, the integral discriminant equals

\begin{equation}
\addtolength{\fboxsep}{5pt}
\boxed{
\begin{gathered}
\ \ \ J_{3|3}\big(S\big) = c_1 \cdot I_{4}^{-1/4} {}_{2}F_{1} \left( \left[ \dfrac{1}{12}, \dfrac{5}{12} \right], \left[ \dfrac{1}{2} \right], - \dfrac{3I_{6}^2}{32I_{4}^3} \right)
+ c_2 \cdot I_{6} I_{4}^{-7/4}  {}_{2}F_{1} \left( \left[ \dfrac{7}{12}, \dfrac{11}{12} \right], \left[ \dfrac{3}{2} \right], - \dfrac{3I_{6}^2}{32I_{4}^3} \right) \ \ \
\end{gathered}
}\label{33}
\end{equation}
\smallskip\\
where $c_{1,2}$ are the two constants, parametrising the general solution of Ward identities. Particular solutions

$$
J^{(1)}_{3|3}\big(S\big) = I_{4}^{-1/4} {}_{2}F_{1} \left( \left[ \dfrac{1}{12}, \dfrac{5}{12} \right], \left[ \dfrac{1}{2} \right], - \dfrac{3I_{6}^2}{32I_{4}^3} \right) \ \ \ \mbox{ and } \ \ \ J^{(2)}_{3|3}\big(S\big) = I_{6} I_{4}^{-7/4}  {}_{2}F_{1} \left( \left[ \dfrac{7}{12}, \dfrac{11}{12} \right], \left[ \dfrac{3}{2} \right], - \dfrac{3I_{6}^2}{32I_{4}^3} \right)
$$
\smallskip\\
are associated with different integration contours and can be called the first and the second branches of $J_{3|3}$. One can see, that the results of this section are parallel to those of sec.3.2 -- even the rational parameters in the hypergeometric function are the same.

\paragraph{Singularities.}Notice, that the point $t = 1$ corresponds to

$$ - \dfrac{3I_{6}^2}{32I_{4}^3} = 1$$
\smallskip\\
which is just the discriminant's vanishing locus $ 32I_{4}^3 + 3I_{6}^2 = D_{3|3} = 0$. To investigate the two other suspicious points $t = 0$ and $t = \infty$, let us study asymptotics at these points. Using (\ref{HG1}), we find

$$
J^{(1)}_{3|3}\big(S\big) \sim I_{4}^{-1/4} + O\big(t\big), \ \ \ \ \mbox{ when } t \rightarrow 0
$$

$$
J^{(2)}_{3|3}\big(S\big) \sim I_{6} I_{4}^{-7/4} + O\big(t\big), \ \ \ \ \mbox{ when } t \rightarrow 0
$$
\smallskip\\
Using (\ref{HG2}), we find

$$
J^{(1)}_{3|3}\big(S\big) \sim \dfrac{- \Gamma(1/2)}{\Gamma(1/12)\Gamma(5/12)} I_{4}^{-1/4} \log \big( 1 + \dfrac{3I_{6}^2}{32I_{4}^3} \big) + O\big(t^0\big), \ \ \ \ \mbox{ when } t \rightarrow 1
$$

$$
J^{(2)}_{3|3}\big(S\big) \sim \dfrac{-\Gamma(3/2)}{\Gamma(7/12)\Gamma(11/12)} I_{6} I_{4}^{-7/4} \log \big( 1 + \dfrac{3I_{6}^2}{32I_{4}^3} \big) + O\big(t^0\big), \ \ \ \ \mbox{ when } t \rightarrow 1
$$
\smallskip\\
Using (\ref{HG1}) and (\ref{HG4}), we find

$$
J^{(1)}_{3|3}\big(S\big) \sim \dfrac{\Gamma(1/2) \Gamma(1/3)}{\Gamma(5/12)^2} \left(\dfrac{3}{32}\right)^{-1/12} I_{6}^{-1/6} + O\big(t^{-5/12}\big), \ \ \ \ \mbox{ when } t \rightarrow \infty
$$

$$
J^{(2)}_{3|3}\big(S\big) \sim \dfrac{\Gamma(1/3) \Gamma(3/2)}{\Gamma(11/12)^2} \left(\dfrac{3}{32}\right)^{-7/12} I_{6}^{-1/6} + O\big(t^{-11/12}\big), \ \ \ \ \mbox{ when } t \rightarrow \infty
$$
\smallskip\\
We can see from these results, that singularities of the integral discriminant $J_{3|3}$ at finite values of $I_{4}, I_{6}$ are completely controlled by the algebraic discriminant $D_{3|3}$. The singularity at discriminant locus is logarithmic. Other singularities, not related to the algebraic discriminant, are situated at infinite values of invariants: $I_{4} \rightarrow \infty$ ($t \rightarrow 0$) and $I_{6} \rightarrow - \infty$ ($t \rightarrow \infty$).

\paragraph{Hypergeometric integral.}Using eq. (\ref{HyperInt}) once again, we obtain

\begin{equation}
\addtolength{\fboxsep}{5pt}
\boxed{
\begin{gathered}
J_{3|3}\big(S\big) = c_1 \cdot \int\limits_{0}^{1} ds \ \dfrac{ s^{-7/12}(1-s)^{-11/12}}{(32 I_4^3 + 3 s I_6^2)^{1/12}} + c_2 I_6 \cdot \int\limits_{0}^{1} ds \ \dfrac{s^{-1/12}(1-s)^{-5/12}}{(32 I_4^3 + 3 s I_6^2)^{7/12}}
\end{gathered}
}\label{integ33}
\end{equation}
\smallskip\\
Just as in the case of $J_{2|4}$, it is possible to make the transformation (\ref{HG3}), which gives

$$
J^{(1)}_{3|3}\big(S\big) = \left( D_{3|3} \right)^{-1/12} {}_{2}F_{1} \left( \left[ \dfrac{1}{12}, \dfrac{1}{12} \right], \left[ \dfrac{1}{2} \right], \dfrac{3I_{6}^2}{32D_{3|3}} \right)
$$

$$
J^{(2)}_{3|3}\big(S\big) = I_{6} \left( D_{3|3}\right)^{-7/12} {}_{2}F_{1} \left( \left[ \dfrac{7}{12}, \dfrac{7}{12} \right], \left[ \dfrac{3}{2} \right], \dfrac{3I_{6}^2}{32D_{3|3}} \right)
$$
\smallskip\\
We conclude, that cases $2|4$ and $3|3$ literally correspond one to another. From the point of view of discriminant theory, this is a remarkable correspondence: discriminant $D_{2|4}$ is a simple two-dimensional discriminant, which is well-known and studied, while $D_{3|3}$ is a three-dimensional discriminant, much more complicated and less widely known. The study of integral discriminants reveals a parallel between these two cases.

\section{Conclusion}

In this paper we have described the first steps into the study of non-Gaussian averages

$$ J_{n|r}\big(S\big) = \int dx_1 \ldots dx_n \ e^{-S(x_1, \ldots, x_n)} = \mbox{ function of invariants of } S$$
\smallskip\\
which are functions only of $SL(n)$ invariants of $S$, because of inherent $SL(n)$ symmetry. The real motivation of this study is generalisation of the well-known Gaussian formula

$$ J_{n|2}\big(S\big) = \int dx_1 \ldots dx_n \ e^{-S_{ij} x_i x_j} = \dfrac{1}{\sqrt{\det S}}$$
\smallskip\\
to non-quadratic forms $S$. If found, such generalisation may immediately have a wide range of applications in statistical physics and quantum field theory. We have worked out several low-dimensional cases and present the results in the form of the following table:

{\fontsize{9pt}{0pt}\[
\begin{array}{c|c|c|c|l}
n & r & \mbox{ Invariants } & \mbox{ Discriminant } D_{n|r} & \mbox{ Integral discriminant } J_{n|r} \\
& & & \\
\hline
& & & \\
2 & 2 & I_2 & I_2 & \ I_{2}^{-1/2} \\
& & & \\
& & & \\
2 & 3 & I_4 & I_2 &\ I_{4}^{-1/6} \\
& & & \\
& & & \\
2 & 4 & I_2, I_3 & I_2^3 - 6 I_3^2 &\ I_{2}^{-1/4} \cdot \sum\limits_{i = 0}^{\infty} \dfrac{1}{i!} \cdot \dfrac{(1/12)_{i} (5/12)_{i} }{(1/2)_{i}} \cdot \left( \dfrac{6I_{3}^2}{I_{2}^3} \right)^i \\
& & & \\
& & & \\
2 & 5 & I_4, I_8, I_{12} & I_4^2 - 64 I_8 & \ I_{4}^{-1/10} \cdot \sum\limits_{i,j = 0}^{\infty} \dfrac{1}{i!j!} \cdot \dfrac{(3/10)_{i + j} (1/10)_{2i + 3j} (1/10)_{j} }{(2/5)_{i + 2j} (3/5)_{i + 2j}} \cdot \left( \dfrac{16I_{8}}{I_{4}^2} \right)^i \left( \dfrac{128I_{12}}{3I_{4}^3} \right)^j  \\
& & & \\
& & & \\
3 & 2 & I_3 & I_3 & \ I_{3}^{-1/2} \\
& & & \\
3 & 3 & I_4, I_6 & 32 I_4^3 + 3 I_6^2 & \ I_{4}^{-1/4} \cdot \sum\limits_{i = 0}^{\infty} \dfrac{1}{i!} \cdot \dfrac{(1/12)_{i} (5/12)_{i} }{(1/2)_{i}} \cdot \left( -\dfrac{3I_{6}^2}{32 I_{4}^3} \right)^i \\
& & & \\
& & & \\
\ldots & \ldots & \ldots & \ldots & \ldots \\
\end{array}
\]}
where

$$
(a)_k \equiv \frac{\Gamma(a+k)}{\Gamma(a)} = a(a+1) \dotsb (a+k-1)
$$
\smallskip\\
As one can see, these series posess a nice structure: they are all hypergeometric, i.e, their coefficients are ratios of $\Gamma$-functions. Despite there are too few examples to make far-going conclusions, we conjecture that

$$ J_{n|r}\big(S\big) = \int dx_1 \ldots dx_n \ e^{-S(x_1, \ldots, x_n)} = \mbox{ hypergeometric function of invariants of } S$$
\smallskip\\
To support this conjecture, it is necessary at least to calculate a few more examples. As follows from the table Fig. 3, the next simplest examples would be $J_{2|6}$, $J_{3|4}$ and $J_{4|3}$. Also, because of hypergeometric nature of integral discriminants, it is interesting to find their $q$-deformation, as well as the CFT representations of these objects, in the spirit of \cite{qhypergeom}. Especially interesting is the interplay between non-trivial $q$-deformation and action-independence of integral discriminants. Of course, most interesting would be generalization
of our results to functional integrals, some independent attempts in this direction have already been made \cite{BP}.

\section*{Acknowledgements}

We are indebted to V.Dolotin and V.Tkachev for stimulating discussions. Our work is partly supported by Russian Federal Nuclear Energy Agency
and the Russian President's Grant of Support for the Scientific Schools NSh-3035.2008.2, by RFBR grant 07-02-00547,
by the joint grants 09-01-92440-CE, 09-02-91005-ANF, 09-02-93105-CNRS and by the NWO project 047.011.2004.026.
The work of Sh.Shakirov is also supported in part by the Moebius Contest Foundation for Young Scientists and by the Dynasty Foundation.

\end{document}